\documentclass[epj]{svjour}
\usepackage{epsfig} 
\usepackage{times}
\begin{document}
%
%
\title{Resonances and final state interactions in the 
reaction ${\bf pp{\to}pK^+\Lambda}$}
\author{A.~Sibirtsev\inst{1}, J.~Haidenbauer\inst{2}, H.-W.~Hammer\inst{1}
and S.~Krewald\inst{2}} 
\institute{Helmholtz-Institut f\"ur Strahlen- und Kernphysik (Theorie), 
Universit\"at Bonn, Nu\ss allee 14-16, D-53115 Bonn, Germany \and
Institut f\"ur Kernphysik (Theorie), Forschungszentrum J\"ulich,
D-52425 J\"ulich, Germany}
\date{Received: date / Revised version: date}

\abstract{
A study of the strangeness production reaction 
$pp{\to}pK^+\Lambda$ for excess energies of 
$\epsilon \le$ 150 MeV, accessible at 
high-luminosity accelerator facilities like COSY,
is presented. 
Methods to analyze the Dalitz plot distribution and angular 
spectra in the Jackson and helicity frames are worked out
and 
suitable observables for extracting information on low lying 
resonances that couple to the $K\Lambda$ system 
and for determining the $\Lambda{p}$ effective-range parameters 
from the final state interaction are identified and discussed.
Furthermore, the chances for identifying the reaction
mechanism of strangeness production are investigated.
}

\authorrunning{A. Sibirtsev et al.} \titlerunning{
The reaction ${\rm pp{\to}pK^+\Lambda}$.}

\maketitle

\section{Introduction}

Strangeness production reactions like $pp{\to}pK^+\Lambda$ are
interesting for various reasons. First of all such reactions
require the creation of a new quark flavour which can occur out of
the vacuum but also from the quark-antiquark sea in the protons. 
Thus, a thorough and dedicated study of the strangeness production
mechanism in those reactions has the potential to ultimately 
deepen our understanding of the internal structure of the baryons.
Furthermore, there are indications that several excited states
of the nucleon decay into the $\Lambda K$ channel. However, 
reliable and quantitative information is rather sparse. 
Investigations of $pp{\to}pK^+\Lambda$ might allow to 
significantly improve the available data base. This concerns specifically 
the $S_{11}(1650)$ and $P_{11}(1710)$ resonances.
Finally, the presence of protons as well as $\Lambda$ hyperons
in the final state opens the possibility to study the interaction
between those baryons, which is still poorly known but whose
knowledge is essential for questions related to the validity of 
the $SU(3)$ flavour symmetry.  

Concerning the mechanism of strangeness production in nucleon-nucleon 
($NN$) collisions one has to concede that is not yet understood 
- although there is a significant 
experimental data base and despite of numerous dedicated 
theoretical investigations. Until recently only data at fairly high energies 
were available. The analysis~\cite{Ferrari1,Yao,Wu,Laget,Deloff,Sibirtsev1}
of those data indicated that different
production mechanisms are compatible with the experimental evidence. 
The data on the $pp{\to}pK^+\Lambda$ reaction cross section and also the
momentum spectra of the final $K$-meson and $\Lambda$-hyperon can be
well reproduced either by $K$-meson or $\pi$-meson exchange models.
Only the large amount of $\Lambda$-hyperon recoil polarization
data~\cite{Bunce,Pondrom,Lundberg,Smith} collected at high energies
from the inclusive $pp{\to}\Lambda{X}$
reaction can be considered as evidence for a $\pi$-meson exchange
dominance~\cite{Soffer,Turbiner1,Turbiner2,Sibirtsev14}.

At high energies where the Regge phenomenology is applicable
the energy dependence of the reaction cross section should
indicate the reaction mechanism. But one has to keep in mind that at those 
energies the energy dependence of the $K$-meson and $\pi$-meson
exchange is almost identical. Indeed, in Regge theory the energy 
dependence of the reaction amplitude is
governed by the exchange trajectory $\alpha{(t)}$ 
via $s^{\alpha{(t})}$, where $s$ is the square of the invariant 
collision energy and $t$ is
the squared four-momentum transferred from the initial nucleon to the 
final nucleon or hyperon, 
for the exchange of a non-strange as well as of a strange meson.
The overall data analysis indicates that the pion exchange trajectory 
amounts to 
$\alpha_\pi(t){=}0.85(t{-}m_\pi^2)$, while the kaon exchange is
given by $\alpha_K(t){=}0.7(t{-}m_K^2)$. Within the Regge theory
the difference between the $\pi$ and $K$ trajectories is only 
due to the mass of the
exchange particles and not by the trajectory intercept at $t{=}0$. 
On the other hand, one can certainly say that 
the data exclude a dominance of the $\rho$-meson exchange, 
whose trajectory is given by $\alpha_\rho(t){=}0.5{+}0.9t$. 
The $K^\ast$ and $K^{\ast\ast}$
exchanges have also large intercepts, 0.5 and 0.35, respectively, and
are likewise not supported by the available data 
for the $pp{\to}pK^+\Lambda$ reaction cross section.

Over the last few years the COSY facility has provided a large 
amount of accurate experimental data on strangeness production 
in $NN$ collisions at low 
energies~\cite{Balewski1,Balewski2,Sewerin,Moskal,Wolke,Kowina1,Kowina2,Rozek,Brandt}.
Theoretical model 
studies~\cite{Tsushima1,Sibirtsev6,Tsushima2,Faldt,Sibirtsev7,Shyam1,Shyam2}
that dealt with those data suggested that the excitation of
resonances in the $K\Lambda$ channel could play an important
if not dominant role for the reaction $pp{\to}pK^+\Lambda$ 
in the near-treshold regime. If this is indeed the case
one has to be cautious in extrapolating from the mechanisms 
that dominate at high energies, {\it i.e.} exchanges by different 
meson trajectories, to what happens at low energies. 
Specifically, it cannot be excluded that, say, vector mesons like
the $\rho$ couple strongly to the resonances in question, 
namely the $S_{11}$(1650) and the $P_{11}$(1710), and therefore play
a decisive role in the strangeness production near threshold.
Also, the investigations so far have made clear that the
interaction between the particles in the final state plays a
role and influences significantly the energy dependence of the 
production cross section in the threshold 
region. Among the 
three possible channels ($K\Lambda$, $K N$, $\Lambda N$) 
it is presumably only the final state interaction (FSI) in
the $\Lambda N$ system which is important 
\cite{Balewski3,Moskal1,Hinterberger,Gasparyan,Hanhart,Gasparyan1}. 
The corresponding 
scattering lengths are not well determined from the few
$\Lambda N$ (and $\Sigma N$) scattering data that exist, but
are expected to be in the order of 1 to 2 
fm~\cite{Holzenkamp,Par1,Rijken,Par3}. 

In this paper we present a study of the strangeness production 
reaction $pp{\to}pK^+\Lambda$ in the energy range 
accessible at the COSY accelerator facility, i.e. for 
excess energies up to $\epsilon\approx$ 150 MeV. 
However, it is not the aim of our work to suggest yet another 
model for that reaction.  Rather we want to embark on a more 
general analysis of this reaction. In view of the complexity
of the situation where neither the production mechanism nor
the final-state interaction are reliably known we 
restrict ourselves to the case of unpolarized experiments. 
Also, we consider only two reaction mechanisms,
namely pion exchange and $K$ meson exchange. However, we want
to emphasize that these mechanisms are understood as being
representatives of a whole class of reaction scenarios 
rather than of the concrete processes. Accordingly, $K$ exchange
represents a scenario where there is strangeness exchange
in the production mechanism and where the elementary
reaction amplitude ($KN\to KN$) is governed by $t$-channel
exchange diagrams, so that its energy dependence is rather
smooth. In particular, there are no resonances involved. 
Pion exchange, on the other hand, stands for a scenario
where no strangeness exchange occurs in the production
mechanism. At the same time the elementary reaction 
amplitude ($\pi N\to K\Lambda$) is dominated by resonance
excitations ($S_{11}(1650)$, $P_{11}(1710)$, $P_{13}(1720)$, ...) 
which implies a strong and characteristic energy dependence. 
In our investigation we will look at the consequences of
these two classes of reaction scenarios for the reaction 
$pp{\to}pK^+\Lambda$ and analyse their signature in 
observables like the total cross section, angular distributions
and the Dalitz plot. 
Thereby we will address the following questions:
(a) Is it possible to discriminate between different production
mechanisms?
(b) What can be learned about the FSI? Is it possible to 
extract the effective-range parameters for the $\Lambda p$
interaction?
(c) Can one determine the parameters of the $S_{11}$(1650) and 
the $P_{11}$(1710) resonance from the strangeness production
reaction?
 
There are, of course, additional important and more concrete
questions. For example, is the strangeness-exchange mechanism
dominated by the $KN\to KN$ reaction or rather by $K^*N\to KN$?
Likewise, are the resonances predominantly excited by pion
exchange or is $\rho$ exchange important as well? Those questions
will not and cannot be addressed in the present investigation. 
For that a throrough investigation of the spin dependence of
the various observables is required which is beyond the scope
of the present paper. 

Our paper is organized as follows: 
In Sect. 2 we introduce those experimental observables which can
be used as a tool for investigating the properties of the $\Lambda p$
final state interaction and of resonances in the $K\Lambda$ channel
but also for a possible identification of the reaction dynamics.
Sect. 3 provides the overall structure of the reaction amplitude and
describes the 
explicit application to the $\pi$ and $K$-meson exchange mechanisms.
The $KN{\to}KN$ and $\pi{N}{\to}K\Lambda$ transition amplitudes
and a description of the treatment of the $\Lambda{p}$ final state 
interaction are given in Sects. 4, 5 and 6, respectively. 
The total $pp{\to}pK^+\Lambda$ reaction 
cross section is analyzed in Sect. 7, while Sects. 8 and 9 
focus on the Dalitz plot and on angular correlations. 
The paper ends with a summary of our results and some concluding 
remarks. 

\begin{figure}[t]
\vspace*{-10mm}
\centerline{\hspace*{8mm}\psfig{file=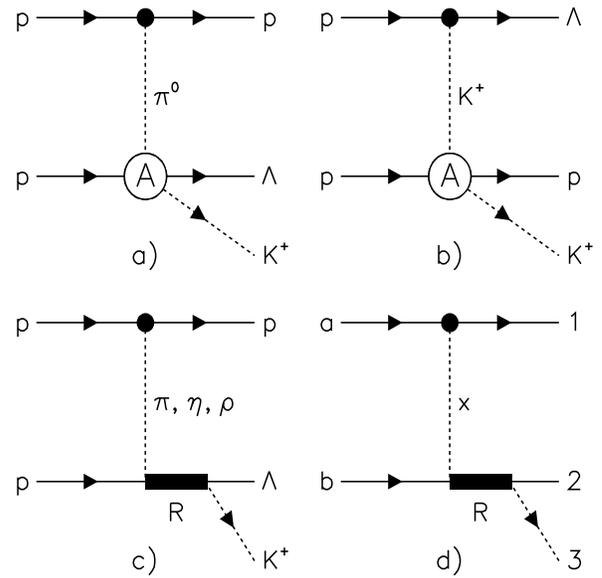,width=9.2cm,height=9.5cm}}
\vspace*{-5mm}
\caption{(a) pion and (b) kaon exchange contributions to 
the reaction $pp{\to}pK^+\Lambda$, included in our investigation. $A$ 
denotes the full ($KN$ or $\pi N \to K^+\Lambda$) transition amplitudes. 
(c) Representation of the resonance model for the reaction $pp{\to}pK^+\Lambda$.
(d) General diagram for $t$-channel contributions to the reaction 
$a+b{\to}1+2+3$ with intermediate resonances $R$ coupled to the $\{23\}$ 
subsystem and exchange particle $x$.}
\label{diaga}
\end{figure}

\section{Kinematic constraints and relevant observables}
In this Sect. we provide a detailed discussion of those observables 
that are directly related to the issues we want to
address (FSI effects, Resonance parameters, production mechanism).
Some of these observables like Dalitz plot distributions or
invariant mass spectra are well known and widely used in the 
analysis of experiments.
Other observables discussed below can be determined only
through the full exclusive reconstruction of the reaction events, 
which can be done only at a some specific experimental facilities like COSY. 
Since the formalism can be applied to any three-body final
state reaction we discuss it in a more general form. Generally speaking
the formalism can be applied to the analysis of any meson 
production in $NN$ collisions ($\pi$, $\eta$, $\omega$, ...),
independently of the collision energy.   

For the consideration of different kinematical variables it is convenient to 
express the invariant amplitude for the reaction $a{+}b{\to}1{+}2{+}3$,
as depicted in Fig.~\ref{diaga}d), in terms of one initial and four
final independent invariants, namely by 
\begin{eqnarray}
s&=&(P_a+P_b)^2 \nonumber \\
s_1&=&(P_1+P_2)^2 \nonumber \\
s_2&=&(P_2+P_3)^2 \nonumber \\
t_1&=&(P_a-P_1)^2 \nonumber \\
t_2&=&(P_b-P_3)^2 \ ,
\label{inv}
\end{eqnarray}
where the $P_i$ denote the four momentum of the corresponding particle $i$. 
The fifth independent final  
variable is the azimuthal rotation angle $\phi$
around the beam axis. We also use the excess energy $\epsilon=\sqrt{s}-m_1^2
-m_2^2-m_3^2$. The differential reaction cross section can then be
written as a function of the four invariants at fixed $s$,
\begin{eqnarray}
\frac{d\sigma}{ds_1 ds_2 dt_1 dt_2}=
\frac{|{\cal M}(s,s_1,s_2,t_1,t_2)|^2}{2^{10}\pi^4 
\lambda(s,m_a^2,m_b^2) \sqrt{-\Delta_4}} \ ,
\label{full}
\end{eqnarray}
where the  \, $\phi$ \, dependence is already \, integrated \, out. \, Here 
$\lambda(x,y,z){=}(x{-}y{-}z)^2{-}4yz$ is the K\"allen function
and $\Delta_4$ is the 
Gramm determinant of a 4$\times$4 symmetric matrix whose elements 
are a combination of $s$, $s_1$, $s_2$, $t_1$, $t_2$ and of the masses 
of the initial and final particles~\cite{Byckling}. 
The physical region of
the invariants is determined by the condition $\Delta_4{\le}0$. The
integration of Eq.~(\ref{full}) over $t_1$ and $t_2$ results in the
famous Dalitz plot~\cite{Dalitz}
\begin{eqnarray}
\frac{d\sigma}{ds_1 ds_2}= \frac{|{\cal M}(s,s_1,s_2)|^2}
{2^8\pi^3 s \lambda^{1/2}(s,m_a^2,m_b^2)} \ .
\label{dalitz}
\end{eqnarray}
If the reaction amplitude is constant, i.e. ${\cal M}{=}const.$, then the 
distribution of the Dalitz plot is isotropic. Therefore, any resonance 
or FSI can be detected through a Dalitz plot analysis. However, 
possible structures in the subsystem formed by particles 1 and 2 
(we use the shorthand notation $\{12\}$ etc. in the following), 
say, can interfere with 
those appearing in the $\{23\}$ subsystem because the differential 
cross section
is a function of both invariants $s_1$ and $s_2$, and this dependence does
not factorize. Such interferences might produce so-called kinematic
reflections in the Dalitz plot projections, {\it i.e.} in the 
invariant mass spectra.
In case of the reaction $pp{\to}pK^+\Lambda$ the Dalitz distribution is a 
useful
tool to study resonances in the $K^+\Lambda$ system and also the
$\Lambda{p}$ (final state) interaction.  

Obviously the Dalitz plot presents already partially integrated data, while
the full information about the reaction dynamics for an unpolarized 
3-body final state is given explicitly by Eq.~(\ref{full}). Different
specific and more practical variables were proposed by Gottfried
and Jackson considering the $\{23\}{\to}2{+}3$ decay and possible
resonances coupled to the $\{23\}$ subsystem~\cite{Gottfried,Jackson}. 
Furthermore, for a more general
case it is interesting to investigate the relation between
the production mechanism and the angular correlations in the decay of
the unstable intermediate particle. In that case it is more useful 
to consider the decay $\{23\}{\to}2{+}3$ and to measure the
angular distribution of particle $3$ in the rest frame of $\{23\}$.

In order to understand the meaning of the angular correlations 
one best considers the reaction 
$a{+}b{\to}1{+}2{+}3$ in terms of the subprocess 
$a{+}b{\to}1{+}\{23\}$ and the subsequent $\{23\}{\to}2{+}3$ decay, 
as depicted in Fig.~\ref{diaga}d). In that case
the 3-body phase space can be expressed in terms of
the 2$\times$2-body phase space convolution as
\begin{eqnarray}
d\Phi_3 = d\Phi_2(s,m_1^2,s_2)\,  d\Phi_2(s_2,m_2^2,m_3^2) \, ds_2,
\end{eqnarray}
where the 2-body phase space, $\Phi_2$, might be taken in different
representations. In particular, it is convenient to use
\begin{eqnarray}
d\Phi_2(s,m_1^2,s_2) &=& \frac{\pi}{2\lambda^{1/2}(s,m_1^2,s_2)}\,
dt_1, \ \ \ \mbox{or} \nonumber \\
d\Phi_2(s_2,m_2^2,m_3^2) &=& \frac{\lambda^{1/2}(s_2,m_2^2,m_3^2)}
{8s_2} \, d\Omega_3.
\label{two}
\end{eqnarray}
Both forms for $d\Phi_2$ are equivalent because of the relation 
between the four momentum transfer and the scattering angle for
the 2-body scattering process. Here $\Omega_3$ is the 
solid scattering angle of particle 3 in the $\{23\}$ rest frame.
In principle, in the $\{23\}$ rest frame the orientation of the momentum 
vector ${\bf p_3}$ can be expressed by the vector
${\bf p_b}$ as well as by ${\bf p_1}$. The first selection corresponds to 
the Jackson frame while the selecting ${\bf p_1}$ corresponds to the
helicity frame. 

Eq.~(\ref{two}) indicates the physical
meaning of the solid angle $\Omega_3$ for the diagrams depicted in
Fig.~\ref{diaga} and naturally defines the axis ${\bf p_b}$ 
along which the angular distribution of particle $3$ should be measured
in the rest frame $\{23\}$. This solid angle is defined as
$\Omega_{3b}$. Indeed, considering the
subprocess $x{+}b{\to}2{+}3$ it is clear that any resonance
structure appearing in the $\{23\}$ subsystem will be directly
reflected in the angular distribution in the Jackson frame.
With respect to $\Omega_{3b}$ the differential cross section is
given by
\begin{eqnarray}
\frac{d\sigma}{ds_2 dt_1 d\Omega_{3b}}{=} 
\frac{\lambda^{1/2}(s_2,m_2^2,m_3^2)}{2^{10} \pi^4
\lambda(s,m_a^2,m_b^2) s_2} |{\cal M}(s,s_2,t_1, \Omega_{3b})|^2.
\label{3dim}
\end{eqnarray}
The polar angle of $\Omega_{3b}$ is called Jackson 
angle $\theta_{3b}$~\cite{Gottfried,Jackson}, while the azimuthal angle
$\phi_{3b}$ was originally defined by Treiman and Yang~\cite{Treiman}. 
The relation between the Jackson angle and the invariants is given by
\begin{eqnarray}
\cos\theta_{3b}{=}
\frac{2s_2(t_2{-}m_b^2{-}m_3^2){+}(s_2{+}m_b^2{-}t_1)
(s_2{+}m_3^2{-}m_2^2)}
{\lambda^{1/2}(s_2,m_b^2,t_1)\, \lambda^{1/2}(s_2,m_3^2,m_2^2)} \ , 
\label{jackson}
\end{eqnarray}
which follows from the definition of $t_2$ in Eq.~(\ref{inv}) when
considering the reaction $x{+}b{\to}2{+}3$.
If the $x{+}b{\to}2{+}3$ amplitude does not depend on  
$\Omega_{3b}$, the angular distribution 
$d\sigma{/}d\Omega_{3b}$ in the Jackson frame is
isotropic. The angular distribution reflects the
$\Omega_{3b}$ dependence of the elementary $x{+}b{\to}2{+}3$ 
reaction amplitude. 
It is important that due to the symmetry of the 
reaction with respect to the beam and target nucleon the Jackson
angle is likewise given along the beam axis, {\it i.e.} particle $b$
can be replaced by $a$ in the previous formulation. 
Moreover, considering
the production of the $\{23\}$ subsystem in a specific spin state one
can parameterize the decay angular distribution  
$d\sigma{/}d\Omega_{3b}$ in terms of the spin density 
matrix~\cite{Jackson,Treiman,Koch}.

The helicity frame defines the solid angle $\Omega_{31}$ with
$\theta_{31}$ being called the helicity polar angle, while $\lambda_{31}$ is
the corresponding azimuthal polar angle. The helicity frame can be naturally 
explained by considering the Dalitz plot representation. 
The solid angle $\Omega_{31}$ 
appears through an extension of the invariant mass $s_3=(P_1{+}P_3)^2$ 
in the frame specified by 
$P_2{+}P_3=(P_a{+}P_b){-}P_3{=}(\sqrt{s_2},\bf{0})$,   
{\it i.e.} in the $\{23\}$ rest frame. 
The helicity polar angle is then given by
\begin{eqnarray}
\cos\theta_{31}{=}\frac{2s_2(m_1^2{+}m_2^2{-}s_1){+}
(s{-}s_2{-}m_1^2)(s_2{+}m_2{-}m_3)}
{\lambda^{1/2}(s,s_2,m_1^2)\lambda^{1/2}(s_2,m_2^2,m_3^2)}
\label{helic}
\end{eqnarray}
and is contained in the Dalitz plot. For fixed $s_2$ the allowed range
of $s_1$ is given by Eq.~(\ref{helic}) with
$\cos\theta_{31}{=}{\pm}1$. Actually Eq.~(\ref{helic}) defines the
contour of the Dalitz plot in the $s_1$ versus $s_2$ plane. 
Any anisotropy in the helicity polar-angle distribution is not necessarily
a signature for the appearance of higher partial waves in the final
system. Rather it reflects structures in the invariant mass
spectra of the $\{12\}$ and $\{23\}$ subsystems. Indeed, for any 
fixed value of $s_2$ it is possible to project the Dalitz plot into
the $s_1$ distribution, which can be converted into a $\theta_{31}$
distribution by Eq.~(\ref{helic}). The same can be done also for 
the $s_2$ projection.

The Chew-Low plot is obtained by integration of
Eq.~(\ref{3dim}) with respect to the solid angle $\Omega_{3b}$ and
yields 
\begin{eqnarray}
\frac{d\sigma}{ds_2 dt_1}{=} 
\frac{\lambda^{1/2}(s_2,m_2^2,m_3^2)}{2^{8} \pi^3
\lambda(s,m_a^2,m_b^2) s_2}\,  |{\cal M}(s,s_2,t_1)|^2 \ ,
\label{chew}
\end{eqnarray}
assuming that the matrix element does not depend on 
$\Omega_{3b}$ (or $\Omega_{31}$). The
Chew-Low presentation is the most convenient way for the evaluation 
of the reaction cross section. It
allows to account for the $t_1$ dependence of the reaction
amplitude via the operator structure of the vertex, the propagator and 
the form factors, as well as of the mass structure in 
only one of the final two-body subsystems. 

\section{The reaction amplitude}
We consider the target $a$ as a spin $1/2$ particle and the exchange 
of a spin-less boson $x$ with mass $\mu$. At this stage we do not account
for the FSI. The most general form of the production amplitude 
is then given by
\begin{eqnarray}
{\cal M}=\frac{f_{a1x}}{\mu}\, F(t_1)\, {\bar u}(p_1){\cal O}u(p_a)\, 
\frac{{\cal A}_{xb{\to}23}(s_2,t_2)} {t_1-\mu^2} \ ,
\label{ampli1}
\end{eqnarray}
where $f_{a1x}$ is the coupling constant of the $a1x$ vertex
and $F(t_1)$ is the form factor at this vertex. The operator 
${\cal O}$ is $\gamma_5$ or $1$ depending on the parity of the
exchanged boson. For $\pi$, $\eta$, $\eta^\prime$, etc. exchanges it is
$\gamma_5$, while for $\sigma$, $a_0$, $f_0$ exchanges it is just
$1$. The quantity ${\cal A}_{xb{\to}23}$ is the invariant amplitude 
for the process $x{+}b{\to}2{+}3$. It is related to the 
physical scattering amplitude and can be 
parametrized through 
\begin{eqnarray}
|{\cal A}_{\pi{N}{\to}K\Lambda}|^2&=&64\pi^2s_{K\Lambda}\, \left[
\frac{\lambda(s_{K\Lambda},m_\pi^2,m_N^2)}
{\lambda(s_{K\Lambda},m_K^2,m_\Lambda^2)}\right]^{1/2} \!\!
\frac{d\sigma}{d\Omega}, 
\nonumber \\
|{\cal A}_{KN{\to}KN}|^2&=&64\pi^2s_{KN} \, \frac{d\sigma}{d\Omega}, 
\label{par1}
\end{eqnarray} 
by utilizing existing differential cross section data for the
two amplitudes in question. In this equation 
$s_{K\Lambda}$ and $s_{KN}$ are the squared invariant energies 
of the $K\Lambda$ or $KN$ subsystems, respectively, 
while $m_N$, $m_K$ and $m_\Lambda$ are the masses of the nucleon, 
the kaon and the $\Lambda$-hyperon.
Since the data determine only the on-shell values of 
${\cal A}_{KN{\to}KN}$ and ${\cal A}_{\pi{N}{\to}K\Lambda}$ 
the off-shellness of the amplitude in Eq.~(\ref{ampli1}) has to
be taken into account. The minimal modification of the on-shell
amplitude to account for this is to include a form factor. 

In any case, 
the $x{+}b{\to}2{+}3$ invariant scattering amplitude can
be expressed in terms of partial waves via~\cite{Hoehler}
\begin{eqnarray}
{\cal A}_{xb{\to}23}{=}8\pi\sqrt{s_2}\,
\chi_f^+ \left[f_1{+}\frac{ ( 
{\mbox{\boldmath $\sigma$}\cdot \bf q}_f) 
( {\mbox{\boldmath $\sigma$}\cdot \bf q}_i)}{q_fq_i}\, f_2
\right]\chi_i,
\label{inva1}
\end{eqnarray} 
where $f_1$ and $f_2$ are defined by 
\begin{eqnarray}
f_1&=&\sum_{l=1}^{\infty} [T^+_{l-1}(s_2)-T^-_{l+1}(s_2)]P_l^\prime(\cos\theta)
\nonumber \\ 
f_2&=&\sum_{l=1}^{\infty}[T^-_l(s_2)-T^+_l(s_2)] P_l^\prime(\cos\theta ) \ .
\label{pwa}
\end{eqnarray}
Here \, $l$ \, is \, the orbital angular momentum of the final state,
$P^\prime_l(\cos\theta)$ is the derivative of the Legendre 
polynomial $P_l(\cos\theta)$, and $\theta$ is
the scattering angle in the $x{+}b{\to}2{+}3$ center-of-mass (cm) system. 
Note that $T^+_l$ and $T^-_l$ are the partial wave (PW) amplitudes 
corresponding to the total angular momentum $J = l{\pm}1/2$. 
In Eq.~(\ref{inva1}) $\chi_i$ and $\chi_f$ are
the two dimensional Pauli spinors of initial and final fermions and
${\bf \sigma}$ are the Pauli spin matrices. 
Furthermore, ${\bf q_i}$ and ${\bf q_f}$ are the cm momenta  
of the initial and final states whose moduli are given by
\begin{eqnarray}
q_i^2{=}\frac{\lambda(s_2,m_x^2,m_b^2)}{4s_2}, \,\,
q_f^2{=}\frac{\lambda(s_2,m_2^2,m_3^2)}{4s_2}
\end{eqnarray}
for on-shell scattering.
The $x{+}b{\to}2{+}3$ differential cross section in terms of the
invariant amplitude is 
\begin{eqnarray}
\frac{d\sigma}{d\Omega}=\frac{1}{64\pi^2s}\, \frac{q_f}{q_i} \,
|{\cal A}_{xb{\to}23}|^2,
\end{eqnarray}
while in terms of the amplitudes 
$F = f_1 + f_2 \cos\theta$ and $G=-f_2\sin\theta$
it is given by 
\begin{eqnarray}
\frac{d\sigma}{d\Omega}{=}
\frac{q_f}{q_i}\,(|F|^2{+}|G|^2) \ .
\label{pwa7a}
\end{eqnarray}
$F$ and $G$ are the non-flip and spin-flip amplitudes,
respectively, and their partial wave representations read
\begin{eqnarray}
F{=}\sum_{l=0}^{\infty}
[(l+1)T_l^+{+}lT_l^-] \, P_l(\cos\theta) \nonumber \\
G{=}\sum_{l=1}^\infty \sin\theta \, 
[T_l^+{-}T_l^-] \, P_l^\prime(\cos\theta).
\label{pwa7}
\end{eqnarray}

For the computation of the $\pi$- and $K$-meson 
exchange mechanisms we need the elementary
$K^+p{\to}K^+p$ and $\pi^0p{\to}K^+\Lambda$ amplitudes
and also the parameters of the corresponding 
pion and kaon emission vertices ($\pi{NN}$ and $\Lambda{NK}$ 
coupling constants and cut-off mass of the pertinent 
vertex form factors), 
cf. diagrams a) and b) in Fig.~\ref{diaga}. 
The elementary $KN{\to}KN$ and $\pi{N}{\to}K\Lambda$ amplitudes 
are specified in the next sections. 
With regard to the couplings $f$ we use the standard relation
to the (pseudoscalar) coupling constants $g$, 
\begin{eqnarray}
f_{\pi{NN}}{=}g_{\pi{NN}}\frac{m_\pi}{2m_N}, \ \  
f_{{\Lambda}NK}{=}g_{{\Lambda}NK}\frac{m_K}{2\sqrt{m_Nm_\Lambda}} \ ,
\end{eqnarray}
and we take the value $g_{{\pi}NN}$=13.45. 
The ${\Lambda{NK}}$ coupling constant is fixed by applying 
standard $SU(3)$ symmetry relations, 
\begin{eqnarray}
g_{\Lambda{NK}}=-g_{\pi{NN}}\frac{1+2\alpha}{\sqrt{3}},
\end{eqnarray}
where $\alpha$ is the $F$ to $D$ ratio, $\alpha{=}F/(F{+}D)$. 
Adopting the quark model estimate of $\alpha$=2/5 together 
with $g_{{\pi}NN}$ specified above we obtain 
$g_{\Lambda{NK}}$=--13.98. 
We furnish the $\pi{NN}$ and $\Lambda{NK}$ vertices with monopole
form factors
\begin{eqnarray} 
F(t_1)=\frac{\Lambda^2_x-\mu^2}{\Lambda^2_x-t_1},
\label{monopole}
\end{eqnarray}
utilizing different cut-off masses for the $\pi$ and $K$-meson
exchanges. These cut-off masses are considered as free
parameters and are adjusted to the data. For COSY energies, the exact
form of this form factor is not important.

In phenomological approaches the relative phase between
the amplitudes is not fixed so that the total reaction 
amplitude is given by 
\begin{eqnarray}
{\cal M}={\cal M}_K +{\cal M}_\pi e^{i\psi} \ ,
\label{phase}
\end{eqnarray}
where $\psi$ can, in principle, depend on the energy. 
The importance of the relative phase or, more generally speaking,
the role of interference effects between the $\pi$ and $K$-meson 
exchange remains so far unclear. 
In the $\pi{+}K$ calculations of
Refs.~\cite{Sibirtsev1,Sibirtsev9} it was found that $K$-meson
exchange dominates the reaction $pp{\to}pK^+\Lambda$ and 
the interference was neglected. 
Later on the role of interference effects was exploited in
Ref.~\cite{Gasparian} in a study of the $pp{\to}pK^+\Lambda$
to $pp{\to}pK^+\Sigma^0$ cross section ratio. 
But also in this work it was concluded that the reaction 
$pp{\to}pK^+\Lambda$ itself is insensitive to the interference, 
because it is dominated by $K$-meson exchange. 
Here we study the $\pi$ and $K$-meson exchanges separately, {\it i.e.} 
we do not add the amplitudes as indicated by Eq.~(\ref{phase}) 
and therefore the uncertainty of the relative phase $\psi$ is 
not relevant for the present investigation. 

\section{The {\boldmath $KN{\to}KN$} amplitude}

\begin{figure}[t]
\vspace*{-6mm}
\centerline{\hspace*{6mm}\psfig{file=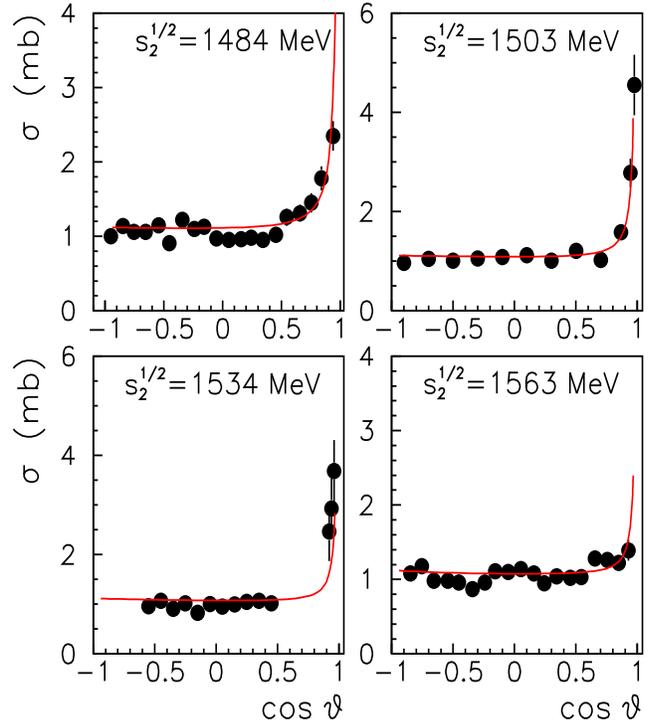,width=9.8cm,height=11.cm}}
\vspace*{-4mm}
\caption{Differential cross sections of the reaction $K^+p{\to}K^+p$ in 
the center of mass system at different invariant collision energies. 
The solid lines show the results from the J\"ulich model \cite{Hoffmann}.
The data are from Ref.~\cite{Knasel}. 
}
\label{erl7}
\end{figure}

We use the $KN$ amplitude of the J\"ulich meson-exchange model. 
A detailed description of the model is given in
Refs. \cite{Buettgen,Hoffmann}. The model yields a satisfactory 
description of the available 
experimental data on elastic and charge exchange $KN$ scattering 
including angular spectra and polarization observables up to 
a $KN$ invariant mass of $\sqrt{s_2}{\simeq}2$~GeV. For the analysis of 
the $pp{\to}pK^+\Lambda$ reaction only the $K^+p{\to}K^+p$
scattering amplitude is necessary. Fig. \ref{erl7} shows the
differential cross section for elastic $K^+p$ scattering at
different invariant energies. The strong forward peaking of the data
and in the calculation comes from the Coulomb interaction. Apart of this 
peaking the angular spectra are isotropic indicating a dominance of
the $s$-wave amplitude in the $K^+p{\to}K^+p$ reaction.

Note that the $KN$ amplitude of the
J\"ulich model was extensively used recently for imposing 
limits of the $\Theta^+$ pentaquark width from data on the 
reaction $K^+d{\to}K^0pp$ \cite{Haidenbauer,Sibirtsev4} and for an 
analysis of the DIANA results~\cite{Dolgolenko} where the $\Theta^+$ 
was observated in $K^+$-meson collisions with $Xe$
nuclei~\cite{Sibirtsev5}. 

For the evaluation of the $K$-exchange contribution to the reaction
$pp{\to}pK^+\Lambda$ one needs the $KN$ amplitude for the energy 
range $m_K{+}m_N{\le}\sqrt{s_2}\le\sqrt{2m_N(2m_N{+}T)}-m_\Lambda$,
where $T$ is the proton beam energy. The energy at COSY is 
limited to $T{\le}$2.88~GeV which means that $\sqrt{s_2}{\le}$1.9~GeV. 
Thus, the energy range for which the $KN$ model of the J\"ulich group 
was designed is sufficient to analyze data in the COSY regime.  
However, in order to connect with data for $pp{\to}pK^+\Lambda$ at 
higher energies, specifically with total cross sections,
one needs to know the $KN$ scattering amplitude at 
$\sqrt{s_2}{>}1.8$~GeV. Here we adopt a phenomenological
approach and parameterize the $KN$ scattering amplitude by 
experimental 
data~\cite{Berestetsky,Ferrari,Sibirtsev2,Sibirtsev3} 
utilizing Eq.~(\ref{par1}).

\section{The {\boldmath $\pi{N}{\to}K\Lambda$} amplitude}

\begin{figure}[t]
\vspace*{-6mm}
\centerline{\hspace*{4mm}\psfig{file=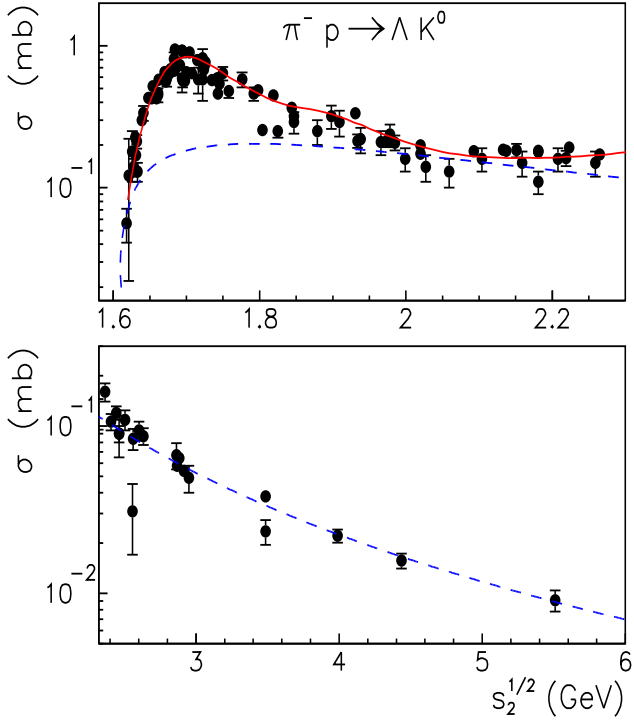,width=9.4cm,height=11cm}}
\vspace*{-4mm}
\caption{Total $\pi^-p{\to}K^0\Lambda$ reaction cross section as a
function of the invariant collision energy. The data are taken from
Ref.~\cite{Landolt}. The solid line is the result based on the 
PW amplitudes of Ref.~\cite{Sotona} while the
dashed line shows the contribution from the $K^\ast$-meson exchange.}
\label{erl5}
\end{figure}

The J\"ulich $\pi N$ model \cite{GaspiN} currently does not include the coupling 
to the $K\Lambda$ channel, therefore we use the PW analysis of Sotona and
Zofka~\cite{Sotona}. Their amplitudes contain ($s$-channel) resonances as well
as $t$-channel $K^\ast$-meson exchange and other background contributions. 
The PW amplitudes 
are given in Ref. \cite{Sotona} up to $\sqrt{s_2}=2.3$~GeV. For 
the analysis of data in the COSY regime we need the $\pi{N}{\to}K\Lambda$ 
amplitude up to $\sqrt{s_2} \le$ 2.05~GeV. At higher 
energies the available $\pi{N}{\to}K\Lambda$ data (differential
cross sections and $\Lambda$-hyperon recoil polarization) can be 
reproduced by $K^\ast$-meson exchange alone taking into account
absorptive corrections~\cite{Krzywicki,Irving1,Hartley,Irving2}.
Thus, we extend the amplitude of Sotona and Zofka appropriately
so that we can study the $pp\to pK^+\Lambda$ reaction cross 
section over a larger energy range and consider data collected at 
COSY as well as those available at higher energies. 
The non-flip and spin-flip amplitudes
for the $K^\ast$-meson exchange is taken from Ref.~\cite{Irving3}
with parameters listed in Ref.~\cite{Sotona}. In order to
reproduce the available data for $\sqrt{s_2}>2.3$~GeV we readjust 
the coupling constants for the $K^\ast$-meson exchange
to $g_0$=--24.0 and $g_1$=--83.3 as compared to those from
Ref.~\cite{Sotona}. 

Fig.~\ref{erl5} shows the total
$\pi^-p{\to}K^0\Lambda$ reaction cross section as a function of
the invariant collision energy~\cite{Landolt}. The solid line is the result
with the PW amplitudes of Ref.~\cite{Sotona}. Obviously, the data 
below 2.3~GeV are fairly well described. The dashed 
line indicates the contribution from the $K^\ast$-meson exchange, which 
dominates the reaction above invariant energies of about 2~GeV.

\begin{figure}[b]
\vspace*{-9mm}
\centerline{\hspace*{5mm}\psfig{file=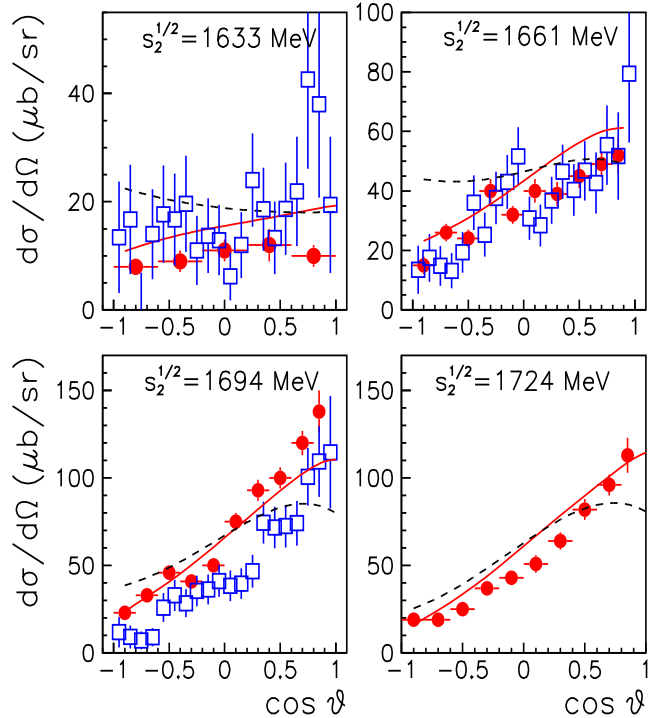,width=9.7cm,height=11.cm}}
\vspace*{-4mm}
\caption{Differential cross sections for the reaction 
$\pi^-p{\to}K^0\Lambda$ in the
center of mass system at different invariant collision energies. 
The solid lines are results based on the full PW amplitudes of 
Ref.~\cite{Sotona}. The dashed lines are obtained when only the
resonant contributions are taken into account.
The squares are data from Ref.~\cite{Knasel}, while the circles 
are from Ref.~\cite{Baker}. 
}
\label{erl4c}
\end{figure}

\begin{figure}[t]
\vspace*{-6mm}
\centerline{\hspace*{7mm}\psfig{file=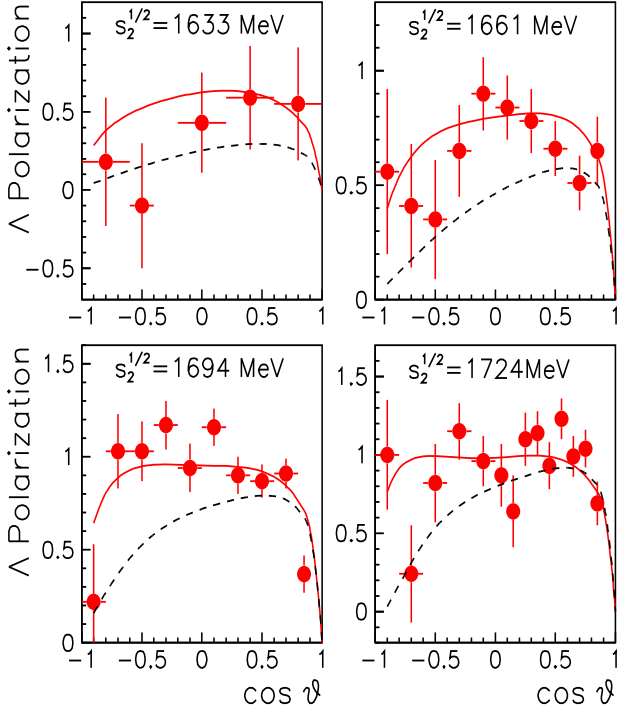,width=9.4cm,height=11.0cm}}
\vspace*{-4mm}
\caption{$\Lambda$ polarizations for the reaction 
$\pi^-p{\to}K^0\Lambda$ in the
center of mass system at different invariant collision energies. 
Same description as in Fig.~\ref{erl4c}
}
\label{erl1}
\end{figure}

A typical feature of the $\pi{N}{\to}K\Lambda$ reaction is the 
strong angular asymmetry and the large $\Lambda$-hyperon 
polarization which occurs already at energies close to the reaction 
threshold. Corresponding experimental results are shown in 
Figs.~\ref{erl4c} and \ref{erl1} where the squares are data from 
Knasel {\it et al.}~\cite{Knasel}, while the circles are from 
the experiment of Baker {\it et al.}~\cite{Baker}. 
Evidently, the polarization is already nonzero at $\sqrt{s_2}{=}1633$~MeV,
the lowest energy where data are available, which corresponds to 
an excess energy of only $\epsilon{=}19.67$~MeV. 
The recoil polarization is defined as
\begin{eqnarray}
P=\frac{2\Im(FG^\ast)}{|F|^2+|G|^2},
\label{recoil}
\end{eqnarray}
where the spin-nonflip ($F$) and spin-flip ($G$) amplitudes are 
given in Eq.~(\ref{pwa7}) in terms of the PW amplitudes. 
The $s$-wave alone results in zero recoil polarization,
while the $p$-wave alone results in a strong angular dependence of 
the polarization. Note that above $\sqrt{s_2}{\simeq}1.8$~GeV
the $\Lambda$-hyperon recoil polarization starts to show a stronger
angle dependence and a change of sign appears at a certain $\cos\theta$.
Let us mention also that $P$ does not vanish even at energies as high as 
$\sqrt{s_2}{=}3.2$~GeV (which is the maximal energy 
where polarization data are available). 

The solid lines in Figs.~\ref{erl4c} and \ref{erl1} are the results 
based on the full PW amplitude of Ref.~\cite{Sotona}, while the 
dashed lines indicate results obtained with inclusion of the resonances 
only. It is clear that the non-resonant background plays a significant 
role already at energies close to the reaction threshold and 
is essential for a quantitative reproduction of the differential
observables. 

One can see from Fig.~\ref{erl4c} that there is partly an 
inconsistency between the two data sets and it is obvious that
the PW analyis cannot reproduce simultaneously both sets 
of data. We should say that there are also polarization data 
by Knasel et al.~\cite{Knasel} for the energy $\sqrt{s_2}{=}1633$~MeV.
However, the error bars of these data are so large that they are not useful 
for our analysis. As a consequence, they are  not shown in Fig.~\ref{erl1}.  

The structure of the elementary $\pi{N}{\to}K\Lambda$ amplitude
suggests that a measurement of the invariant $K\Lambda$ mass spectra and
the angular correlations according to Eqs.~(\ref{jackson}), 
and (\ref{helic}) might allow to isolate
the contribution of the $\pi$-exchange to the
$pp\to pK^+\Lambda$ reaction. However, it is possible that due to
the short-ranged nature of the strangeness production reaction 
higher partial waves in the virtual $\pi{N}{\to}K\Lambda$ amplitude 
are suppresed so that the $s$ wave dominates the 
$pp{\to}pK^+\Lambda$ observables at COSY energies. In that case the
angular correlation of Eq.~(\ref{jackson}) from $\pi$ and $K$-meson
exchanges would be
similar and the relevant spectrum should be isotropic. But it should
be still possible to detect any $s$-wave resonance in the $K\Lambda$ system 
through an analysis of the Dalitz plot distribution and the
angular correlation of Eq.~(\ref{helic}). We will come back to this
issue below. 

\section{ The {\boldmath $\Lambda p$} final state interaction}

Production reactions like $NN\to K^+\Lambda N$ require a large
momentum transfer between the initial and final baryons. Thus,
the range of the production mechanism will be much smaller
than the characteristic range of the interactions in the final
states. In such a case 
the energy dependence of the reaction amplitude is driven primarily by
that of the scattering amplitude of the outgoing particles 
and it was proposed~\cite{Goldberger} to factorize the reaction amplitude
\begin{eqnarray}
{\cal M} \to {\cal M}\times {\cal A}_{FSI},
\label{tore}
\end{eqnarray}
where ${\cal A}_{FSI}$ denotes the amplitude due to the interaction
between the final particles. ${\cal A}_{FSI}$ is in principle 
a 3-body amplitude. However, it is generally assumed that 
the $\Lambda p$ interaction dominates over the other 
possible final-state interactions and therefore one replaces 
${\cal A}_{FSI}$ by ${\cal A}_{\Lambda p}$. 
The validity of this assumption is to some extent questionable. 
It is based primarily on the observation that the absolute value 
of the $\Lambda p$ scattering length is substantially
larger than those for $K^+p$ and $K^+\Lambda$ scattering, although 
one has to admit that the latter is actually not known. 
In any case, very close to the reaction threshold the relative
momenta between all final particles are small and one should account
for the interference between the FSI in the various two-body 
systems. In that kinematics the
interference term between the large and small scattering lengths
might be not negligible. For instance in the
analysis~\cite{Sibirtsev10,Sibirtsev11} of the $\gamma{d}{\to}pn\eta$
reaction very close to the reaction threshold, {\it i.e.} at
$\epsilon{<}$20~MeV it was found that the $NN$ and 
$\eta N$ final state interactions interfere.
But in the present investigation we concentrate on excess energies
in the order of 100 MeV and, therefore, the simplification in the
FSI treatment should be justified.  

According to the above arguments the near threshold mass dependence of
the $\Lambda p$ spectrum for the $pp{\to}pK^+\Lambda$ reaction 
might be dominated by the energy dependence of the $\Lambda p$ scattering 
amplitude. Since 
the range of the $\Lambda p$ invariant mass is from $m_\Lambda{+}m_N$ to
$\epsilon{+}m_\Lambda{+}m_N$ FSI effects should be visible in 
differential observables at any collision energy. On the other hand,
in case of the total reaction cross section FSI effects should be
dominantly seen at energies close to the reaction threshold. 
At higher energies the FSI affects only a small part of the available 
phase space~\cite{Sibirtsev12,Sibirtsev13},
{\it i.e.} only the region where the relative momenta of the $\Lambda p$ 
system are sufficiently small and, therefore, have a comparably low 
weight in the integration over the whole phase space.
 
A very simple treatment of FSI effects was proposed by
Watson~\cite{Watson} and Migdal~\cite{Migdal}. Close to the reaction
threshold the invariant scattering amplitude is dominated
by the $s$ wave and can be expressed in terms of the effective range
expansion as
\begin{eqnarray}
{\cal A}_{\Lambda p}(q)=N_0(m_\Lambda{+}m_N)
\left[-\frac{1}{a}+\frac{rq^2}{2}-iq\right]^{-1},
\label{watson}
\end{eqnarray}
where $a$ and $r$ are the scattering length and the effective range,
respectively, and $q$ is the relative momentum between the $\Lambda$
hyperon and final proton, 
\begin{eqnarray}
q= \frac{\lambda^{1/2}(s_{\Lambda p},m^2_\Lambda,m^2_p)}
{2\sqrt{s_{\Lambda p}}} \ .
\end{eqnarray}
$N_0$ is a normalization constant, which can not be fixed 
within the Watson-Migdal approximation. 

\begin{figure}[t]
\vspace*{-5mm}
\centerline{\hspace*{3mm}\psfig{file=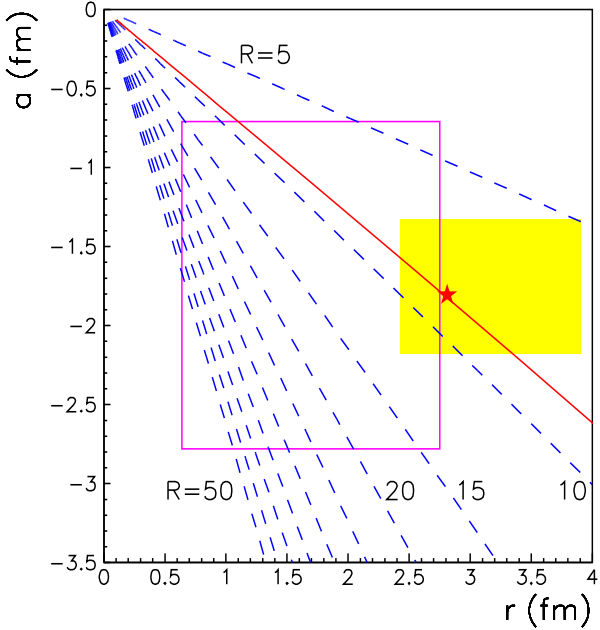,width=8.9cm,height=9.5cm}}
\vspace*{-4mm}
\caption{The enhancement factor $R$ (cf. Eq.~(\ref{enh})) as a
function of the $\Lambda{p}$ effective range $r$ and scattering length
$a$. The dashed lines show results for $R$=5$\div$50 calculated
at $\epsilon$=2~MeV. 
The hatched (open) box indicates
the range of $r$ and $a$ for the $\Lambda{p}$ interaction in the triplet
(singlet) state, taken from Refs.~\cite{Par1,Rijken,Par3}. 
The asterisk indicates the parameters used in our analysis, while the 
solid line shows the family of $r$ and $a$ resulting in $R$=8.7.}
\label{erka8}
\end{figure}

While the Watson-Migdal prescription is well applicable 
to final-state interactions that are characterized by a large
scattering length like in case of the $NN$ interaction, say, this is not
true for $\Lambda p$ where the expected scattering lengths
are only in the order of one to two fermi \cite{Gasparyan1}.
Here one should resort at least to the so-called Jost function approach
which was found in Ref. \cite{Gasparyan1} to yield resonable qualitative
results. For a $\Lambda p$ scattering amplitude that is given by the 
effective range approximation, cf. Eq.~(\ref{watson}), over the whole
energy range, the FSI factor in the Jost function approach can
be evaluated analytically and amounts to 
\begin{eqnarray}
{\cal A}_{\Lambda p}(q)= \frac{q+i\beta}{q-i\alpha},
\label{jost1}
\end{eqnarray}
where $\alpha$ and $\beta$ are related to the scattering
parameters via
\begin{eqnarray}
a{=}\frac{\alpha+\beta}{\alpha\beta}, \,\,\,\,\,\,
r{=}\frac{2}{\alpha+\beta}
\end{eqnarray}
with $\alpha{<}0$ and $\beta{>}0$. In our notation the scattering
length $a$ is defined with a negative sign, cf. Eq.~(\ref{watson}), 
which explains the difference to the formulas
given by Goldberger and Watson~\cite{Goldberger}. 
Eq.~(\ref{jost1}) implies the limits
\begin{eqnarray}
\lim_{q\to 0} {\cal A}_{\Lambda p}(q){=}-\frac{\beta}{\alpha}, \, \, \,
\, \,\,\lim_{q\to\infty} {\cal A}_{\Lambda p}(q){=}1,
\label{jostlimit}
\end{eqnarray}
which can be used as a measure for the relative strength of
the FSI with respect to the contribution from the processes without
FSI. Eq.~(\ref{jost1}) can be written in the form
\begin{eqnarray}
{\cal A}_{\Lambda p}(q){=}\left[\frac{r\beta^2}{2}{+}
\frac{rq^2}{2}\right]
\left[-\frac{1}{a}+\frac{rq^2}{2}-iq\right]^{-1}\,,
\end{eqnarray}
which at small $q$ is close to Watson-Migdal parameterization of
Eq.~(\ref{watson}) (apart from the unknown normalization constant $N_0$).
In addition, the Jost function approach also includes the correct 
behavior for large momenta, cf. Eq.~(\ref{jostlimit}).

At present a solid estimation of FSI effects for the reaction $pp{\to}pK^+\Lambda$
is difficult because of two reasons: 
(a) The $\Lambda{p}$ system can be in the singlet and triplet states that
can have different effective-range parameters $a$ and $r$. It is not known 
whether the 
$\Lambda{p}$ system is predominantly produced in one or the other state. 
We should mention though that most microscopic models of the reaction 
$pp{\to}pK^+\Lambda$ predict a dominance of the triplet contribution.
(b) The effective range parameters are not well known, i.e. 
they are afflicted with large uncertainties. This is visualized in 
Fig. \ref{erka8} where the hatched box 
shows the range of $r$ and $a$, for the triplet case, taken from
some recent $YN$ potential models ~\cite{Par1,Rijken,Par3}. 
The open box in Fig.~\ref{erka8}
indicates the variation in the singlet effective-range parameters.
It is clear that the uncertainties of the $\Lambda{p}$ interaction 
allow a large freedom of FSI effects in the reaction $pp{\to}pK^+\Lambda$. 

In order to illustrate how strongly the FSI with different scattering
parameters might influence the $pp{\to}pK^+\Lambda$ reaction cross
section we evaluate the so-called enhancement factor $R$ as a function 
of the excess energy. $R$ is defined as the integral
of $|{\cal A}_{\Lambda p}(q)|^2$ from Eq.~(\ref{jost1}) over the 
nonrelativistic 3-body phase space, normalized to the phase space volume
$\Phi_3$, i.e. 
\begin{eqnarray}  
R(\epsilon)&=&\frac{1}{\Phi_3}\!\!\int\limits_o^{\sqrt{2\mu\epsilon}}\!\!
\sqrt{2{\tilde\mu}(\epsilon{-}\frac{q^2}{2\mu})} \, \, 
\frac{q^2+\beta^2}
{q^2+\alpha^2} \, \, q^2 \, dq  \nonumber \\
&=&1+\frac{4\beta^2-4\alpha^2}{(-\alpha+\sqrt{\alpha^2+2\mu\epsilon})^2},
\label{enh}
\end{eqnarray}
where $\mu$ and ${\tilde\mu}$ are reduced masses given by
\begin{eqnarray}  
\mu{=}\frac{m_\Lambda m_N}{m_\Lambda{+}m_N}, \,\,\,\,\,
{\tilde\mu}{=}\frac{m_K(m_\Lambda{+}m_N)}{m_K{+}m_\Lambda{+}m_N},
\end{eqnarray}
and $\Phi_3$ is given by the integral of Eq.~(\ref{enh}) without the
factor $|{\cal A}(q)|^2$ from Eq.~(\ref{jost1}).

The dashed lines in Fig.~\ref{erka8} show the enhancement factor $R$ for the
specific excess energy $\epsilon$=2~MeV, as a function of the effective range 
and scattering length. One can see that the variations of the 
$\Lambda{p}$ triplet parameters in Refs.~\cite{Par1,Rijken,Par3} 
exclude any values $R{>}$12. On the other hand, the
singlet parameters allow for almost any magnitude of the enhancement factor. 
As just mentioned above, most microscopic models of the reaction 
$pp{\to}pK^+\Lambda$ favour the triplet contribution.
In our analysis we do not consider singlet and triplet 
$\Lambda{p}$ FSI effects separately but use averaged parameters fixed
to the value shown in Fig.~\ref{erka8} by the asterisk~\cite{Holzenkamp}.

\section{The {\boldmath $pp{\to}pK^+\Lambda$} reaction cross section}

\begin{figure}[t]
\vspace*{-6mm}
\centerline{\hspace*{3mm}\psfig{file=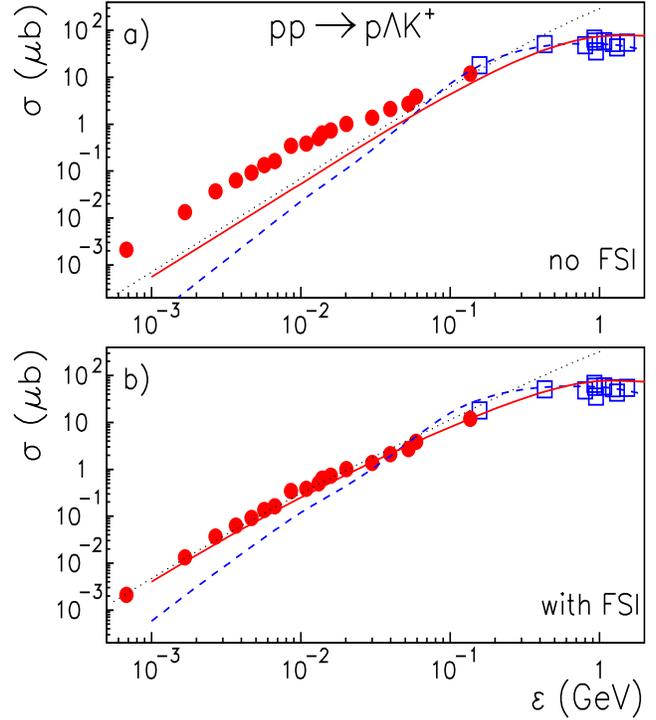,width=9.6cm,height=11.cm}}
\vspace*{-4mm}
\caption{
Total cross section for the reaction $pp{\to}pK^+\Lambda$ as 
a function of the excess energy. 
The upper figure shows results without FSI, while for the lower figure
the $\Lambda{p}$ FSI was included via Eq.~(\ref{jost1}). 
The solid lines are results for the $K$-meson exchange mechanism. 
The dashed lines are obtained with $\pi$-meson exchange and with 
the full $\pi{N}{\to}K\Lambda$ transition amplitude. 
The dotted lines show results with a constant reaction amplitude.
The squares are data taken from Ref.~\cite{Landolt}, while the circles are 
from experiments at the COSY facility~\cite{Balewski2,Sewerin,Kowina2,Brandt}. 
}
\label{erka1}
\end{figure}

Fig.~\ref{erka1} shows the $pp{\to}pK^+\Lambda$ reaction cross section
as a function of the excess energy. The squares represent data that were 
available before the COSY aera, collected in Ref.~\cite{Landolt}. 
The circles are from measurements at the COSY facility, 
performed by the COSY-11~\cite{Balewski2,Sewerin,Kowina2} and TOF
Collaborations~\cite{Brandt}. Apparently the COSY experiments 
provide a substantial contribution to the data base, 
specifically they are the only source of information for the
behaviour of the cross section near the reaction threshold. 

For a general overview it is always useful to compare the data
to the phase space behaviour, {\it i.e.} to consider the given
reaction kinematics but set the reaction amplitude to ${\cal
M}{=}const$. In case of the total reaction cross section the relevant
kinematics is the dependence of the 3-body phase space on the excess
energy, which in the non-relativistic case\footnote{Actually the 
nonrelativistic and the 
relativistic phase space for the reaction $pp{\to}pK^+\Lambda$ are
almost identical at $\epsilon{<}2$~GeV.} is given by Eq.~(\ref{enh}). 
The integration can be performed analytically and yields
\begin{eqnarray} 
\sigma (\epsilon){=}\frac{\sqrt{m_K m_N m_\Lambda}}{2^7 \pi^2 
(m_K{+}m_N{+}m_\Lambda)^{3/2}} \frac{\epsilon^2}{\sqrt{s^2{-}4sm_N^2}}
\, |{\cal M}|^2 \ . 
\label{cross1}
\end{eqnarray}
This results 
is shown by the dotted line in Fig.~\ref{erka1}a) for
the squared invariant amplitude $|{\cal M}|^2{=}2.2{\cdot}10^7$ $\mu$b,
which was normalized to the data at $\epsilon{\simeq}$130~MeV. 
Following the discussion given in Sect.~6
one expects that close to the threshold the data deviate from
a calculation that neglect the $\Lambda{p}$ FSI, and this is indeed 
the case, cf. Fig.~\ref{erka1}a). For example, at $\epsilon{\simeq}$2~MeV
the phase space line underestimates the data by a factor of around 9.

When we now introduce FSI effects within the Jost function 
approach Eq.~(\ref{jost1}) we can easily reproduce the energy
dependence of the data by adopting the parameters
\begin{eqnarray}
\beta{=}212.7~{\rm MeV \, \, \, and\, \, \, \,}
\alpha{=}-72.3~{\rm MeV}, 
\label{param}
\end{eqnarray}
which correspond to the low-energy parameters $a$=--1.8~fm and 
$r$=2.8 fm. The resulting cross
section is shown by the dotted line in Fig.\ref{erka1}b).

The employed low-energy parameters are indicated in Fig.~\ref{erka8} by 
an asterisk. Obviously, they are well within the present 
uncertainty range of the triplet parameters. But we would like to
emphasize that any (singlet- or triplet) combination of effective-range
parameters that lies on the solid line of Fig.~\ref{erka8} would give
similar results, i.e. would reproduce the energy dependence observed in the
experiment. There is no unique solution. 
Thus, the presented specific fit does not provide any deeper 
understanding of the strangeness production mechanism
or the hyperon-nucleon interaction. It only illustrates that any
reaction mechanism which implies a sufficiently weak energy dependence
would be compatible with the empirical information. 

\begin{figure}[b]
\vspace*{-4mm}
\centerline{\hspace*{5mm}\psfig{file=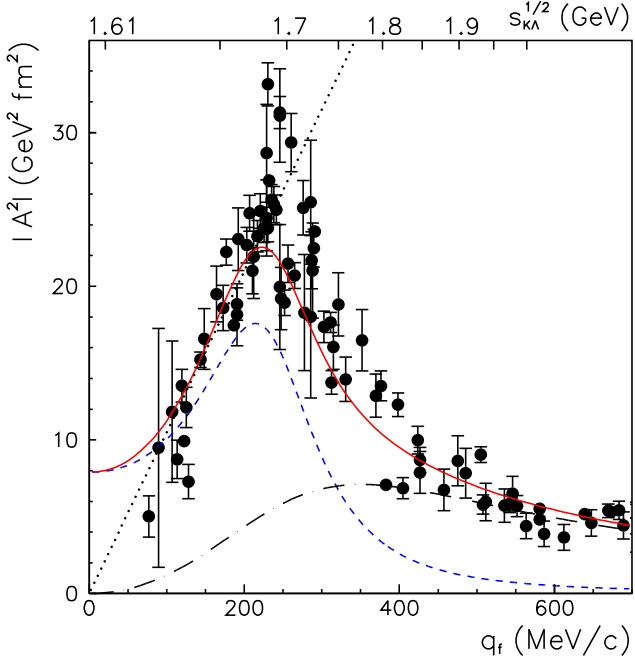,width=9.5cm,height=9.5cm}}
\vspace*{-4mm}
\caption{The $\pi{N}{\to}K\Lambda$ amplitude squared as a function of
the final momentum $q_f$ and the invariant collision energy
$s_{K\Lambda}$ (axis at the top). 
The dashed and dashed-dotted lines show the contribution 
from the $S_{11}$ and $P_{11}$ resonances, respectively, of the
PW analysis presented in Ref.~\cite{Sotona}. 
Their sum corresponds to the solid line.
The dotted line indicates the $q_f$ dependence.
The circles are experimental results, extracted from the data given 
in Ref.~\cite{Landolt}, cf. text. 
}
\label{erka3}
\end{figure}

Let us now come to concrete reaction mechanisms.
The solid line in Fig.~\ref{erka1}a) shows the result for 
$K$-meson exchange without FSI. In our calculation the $s_{KN}$ and
$t_2$ dependence of the $KN{\to}KN$ scattering amplitude, the $t_1$
dependence of the propagator and the $KN\Lambda$ vertex and formfactor and
the $s_{\Lambda{p}}$ dependence of the FSI (cf. below) is taken into
account and we perform a full four-dimensional integration of
Eq.~(\ref{full}). The results were normalized to the data 
at $\epsilon{\simeq}$1~GeV by adjusting the
cut-off mass of the form factor Eq.~(\ref{monopole}) to
$\Lambda{\simeq}$1.7~GeV. In contrast to the pure phase
space, the $K$-meson exchange well reproduces the energy dependence of the 
data also at high energies, which will be discussed later. The results 
obtained with the
FSI of Eq.~(\ref{jost1}) utilizing the parameters of Eq.~(\ref{param})
are shown by the solid line in Fig. \ref{erka1}b).
It is interesting to see that the results for $K$-meson exchange are
practically identical to the phase-space behaviour over a large
energy range.

Results for the $\pi$-exchange mechanism are shown by the dashed line in 
Fig.~\ref{erka1}a), for the case without $\Lambda{p}$ FSI. 
Again we normalize our results at $\epsilon{\simeq}$1~GeV by
adjusting the cut-off mass to $\Lambda{\simeq}$1.6~GeV. It is evident that
the energy dependence predicted by the $\pi$-meson exchange differs 
from the one resulting from $K$-meson exchange and the phase
space calculations. As a consequence, the calculation with FSI 
(dashed line in Fig.~\ref{erka1}b)), substantially underestimates the 
data below $\epsilon$=200~MeV.

In order to shed light on the difference in the energy dependence of the 
total cross sections resulting from $K$ and $\pi$-meson exchange 
let us take a look at the elementary $\pi{N}{\to}K\Lambda$ amplitude
${\cal A}_{\pi{N}{\to}K\Lambda}$. 
The square of this amplitude can be obtained from data via Eq.~(\ref{par1}). 
It is shown in Fig.~\ref{erka3}. Here the angular dependence is integrated
out so that the amplitude depends only on the invariant collision 
energy $s_{K\Lambda}$ or the final momentum $q_f$, respectively. 
The experimental results (solid circles) are cross section data
taken from Ref.~\cite{Landolt}, divided appropriately by phase-space 
factors. It is evident that $|{\cal A}_{\pi{N}{\to}K\Lambda}|^2$ is 
strongly energy dependent. Specifically, it does not exhibit the 
behaviour of a standard $s$-wave amplitude, which 
would be constant in the near-threshold region, nor that of a
$p$-wave, which should be proportional to $q_f^2$. 
Rather the data seem to rise linearly with the moment $q_f$, 
cf. the dotted line in Fig.~\ref{erka3}.

\begin{figure}[t]
\vspace*{-3mm}
\centerline{\hspace*{3mm}\psfig{file=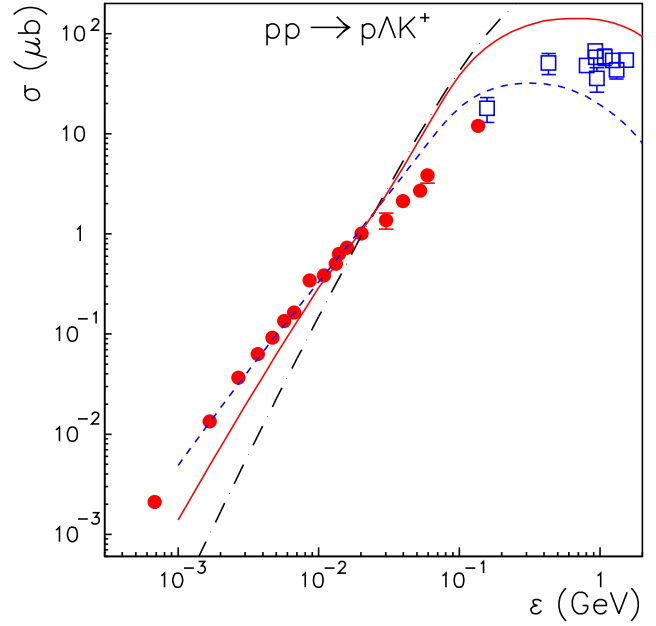,width=9.6cm,height=9.5cm}}
\vspace*{-4mm}
\caption{
Total cross section for the reaction $pp{\to}pK^+\Lambda$ as 
a function of the excess energy. 
The dashed line is the result for the $\pi$-meson exchange mechanism 
with the $S_{11}$(1650) resonance amplitude, while the dashed-dotted line 
was obtained with the $P_{11}(1710)$ resonance alone. The solid line
shows the full calculation. In all cases the $\Lambda{p}$ FSI is taken
into account via Eq.~(\ref{jost1}) with parameters 
specified in Eq.~(\ref{param}). For comparison the lines are
normalized at the same excess energy. 
The squares are data taken from Ref.~\cite{Landolt}, while the circles are 
from experiments at the COSY facility~\cite{Balewski2,Sewerin,Kowina2,Brandt}. 
}
\label{erka2}
\end{figure}

According to the PW analysis of Ref.~\cite{Sotona} the reaction 
$\pi{N}{\to}K\Lambda$ is dominated by the $S_{11}(1650)$ and $P_{11}(1710)$ 
resonances for energies up to $\sqrt{s_{K\Lambda}}{\simeq}$1.8 GeV, 
as discussed in Sect. 5. 
These resonance amplitudes,
${\cal A}^R_{\pi{N}{\to}K\Lambda}$, are given by~\cite{Sotona}
\begin{eqnarray}
{\cal A}^R_{\pi{N}{\to}K\Lambda}{=}{-}\frac{8\pi\sqrt{s_{K\Lambda}}}{q_iq_f}\,
\frac{\sqrt{\Gamma_{\pi{N}}\Gamma_{K\Lambda}}\, M_R\Gamma}
{M_R^2{-}s_{K\Lambda}{-}ifM_R\Gamma},
\label{reson}
\end{eqnarray}
where $M_R$ and $\Gamma$ are the mass and full width of the resonance,
\begin{eqnarray}
f{=}\frac{\alpha}{100}f^l_f{+}\frac{100{-}\alpha}{100}f^l_i, \nonumber
\\ f^l_i{=}\frac{\phi_l(R \, q_i)\,\, q_i}
{\phi_l(R\, q_i^R)\,\, q_i^R}, \,\,\,\,
f^l_f{=}\frac{\phi_l(R\, q_f)\,\, q_f}
{\phi_l(R\, q_f^R)\,\, q_f^R},
\end{eqnarray}
and the initial and final momenta are 
\begin{eqnarray}
q_i^2{=}\frac{\lambda(s_{K\Lambda},m_\pi^2,m_N^2)}{4s_{K\Lambda}}, \,\,
q_f^2{=}\frac{\lambda(s_{K\Lambda},m_K^2,m_\Lambda^2)}{4s_{K\Lambda}} \ .
\end{eqnarray}
$q_i^R$ and  $q_f^R$ are the corresponding momenta at the resonance pole 
position, {\it i.e.} at $\sqrt{s_{K\Lambda}}{=}M_R$. The interaction radius was
taken as $R$=1.696 GeV$^{-1}$. The function $\phi_l$ ensures the correct
threshold energy dependence and is given by~\cite{Baker,Saxon,Bell}
\begin{eqnarray}
\phi_0(x){=}1, \,\,\,\,\, \phi_1(x){=}\frac{x^2}{1{+}x^2} \ , 
\end{eqnarray}
for $s$ and $p$ waves, respectively. 
Finally the partial decay width was parametrized by 
\begin{eqnarray}
\sqrt{\Gamma_{\pi{N}}\Gamma_{K\Lambda}}= B \sqrt{f_f^l f_i^l} \ .
\end{eqnarray}
In the following calculations we use the $S_{11}$ and $P_{11}$
resonance parameters as fixed by the PW analysis of 
Ref.~\cite{Sotona} which we already introduced in the Sect. 5. 
Specifically, we use for the $S_{11}(1650)$ resonance 
\begin{eqnarray}
M_R{=}1678~{\rm MeV},\,\, \Gamma{=}117~{\rm MeV},\nonumber \\
B{=}0.2175, \,\,\, \alpha{=}7.8855,
\label{s11}
\end{eqnarray}
and for the $P_{11}(1710)$ resonance 
\begin{eqnarray}
M_R{=}1730~{\rm MeV},\,\, \Gamma{=}543~{\rm MeV},\nonumber \\
B{=}0.1565, \,\,\, \alpha{=}12.893.
\end{eqnarray}
The square of these resonance amplitudes are shown
in Fig.~\ref{erka3} by the dashed ($S_{11}$) and dash-dotted 
lines ($P_{11}$), respectively. The solid line is the sum of
these two contributions which illustrates that those two
resonances together indeed reproduce the bulk of the 
experimental amplitude. 

Predictions for the $pp{\to}pK^+\Lambda$ cross section utilizing the pion
exchange mechanism with the $S_{11}$ or $P_{11}$ resonances are 
shown in Fig.~\ref{erka2} by the solid and dashed lines, respectively. 
The $\Lambda{p}$ FSI is included via Eq.~(\ref{jost1}) 
with the parameters specified in Eq.~(\ref{param}). It is obvious
that the energy dependence of the calculation based on the $P_{11}$ 
resonance differs substantially from the experiment. 
The curve obtained for the $S_{11}$ resonance is in good agreement with 
the data for $\epsilon{<}$40~MeV, but deviates at higher energies. 

The results discussed above make clear that, 
in contrast to the $K$-meson exchange scenario, 
the pion exchange mechanism yields a much stronger
energy dependence of the production cross section,
due to the excitation of resonances.
However, it would be premature to see the 
individual disagreement of the $S_{11}$ as well as of the $P_{11}$ 
case with the energy dependence of the data as an evidence for a
$K$-meson exchange dominance of the $pp{\to}pK^+\Lambda$ reaction. 
Indeed, by exploiting the 
freedom in the interplay between the $S_{11}$(1650) and $P_{11}(1710)$
resonances it is still possible to reproduce the cross section data
over a large energy range, as is well illustrated in 
Refs.~\cite{Tsushima1,Sibirtsev6,Tsushima2}. To discern between the
two scenarios considered here ($K$ versus $\pi$ exchange) one
must consider differential observables like those introduced in Sect. 2. 
Corresponding results will be discussed in the next two sections. 

Before that we want to comment on the $t_1$ dependence. 
For that purpose we consider the
Chew-Low integration of Eq.~(\ref{chew}) with the reaction amplitude ${\cal
M_\pi}$ neglecting the FSI, {\it i.e.} the $s_{\Lambda{p}}$ dependence. 
After integrating over $t_2$ or $\cos\theta_{3b}$ (see Fig.~\ref{diaga} and 
Eq.~(\ref{jackson})) the $pp\to pK^+\Lambda$ reaction cross section due to 
$\pi$-meson exchange is given by 
\begin{eqnarray}
\sigma(\epsilon)\!&=&\!\frac{g_{{\pi}NN}^2}
{2^8\pi^2(s^2{-}2sm_N^2)}
\int\limits_{s_-}^{s_+}\!\!
ds_{K\Lambda}\!
\int\limits_{t_-}^{t_+}\!\!dt_1\frac{\lambda^{1/2}
(s_{K\Lambda},m_K^2,m_\Lambda^2)}{s_{K\Lambda}}\nonumber \\
&\times&\frac{-t_1}{(t_1{-}m_\pi^2)^2}\,
\left[\frac{\Lambda^2{-}m_\pi^2}
{\Lambda^2{-}t_1}\right]^2
|{\cal A}_{\pi{N}{\to}K\Lambda}(s_{K\Lambda})|^2,
\label{sigma1}
\end{eqnarray} 
where $t_1$ is the squared four-momentum transferred
from the initial to the final proton and the limits of integrations are 
\begin{eqnarray}
s_-&=&(m_K{+}m_\Lambda)^2, \,\,\,\,
s_+{=}(m_K{+}m_\Lambda{+}\epsilon)^2, \nonumber \\
t_\pm&=&2m_N^2{-}\frac{s{+}s_{K\Lambda}{-}m_N^2}{2} \nonumber \\
&\pm&\frac{\sqrt{s{-}4m_N^2}\,\,\,
\lambda^{1/2}(s,s_{K\Lambda},m_N^2)}{2\sqrt{s}}.
\label{range}
\end{eqnarray} 
 
For cut-off masses in the order of 
$\Lambda$=1.6~GeV the $t_1$ dependence of Eq.~(\ref{sigma1}) becomes
signifiant only for $t_1{>}-$0.3 GeV$^2$, which is accessible only at
$\epsilon{\ge}$200 MeV. Indeed, at threshold 
\begin{eqnarray}
t_\pm = m_N(m_N-m_K-m_\Lambda)\simeq -0.63~{\rm GeV}^2,
\end{eqnarray} 
so that for energies not too far from the threshold 
the reaction cross section depends only very weakly on $t_1$. 
Therefore,
for pion exchange -- but in fact, also for kaon exchange --
the $t_1$ dependence of the reaction amplitude ${\cal M}$ is almost negligible 
for excess energies $\epsilon{<}$200 MeV. 
Only for energies around $\epsilon\approx$1 GeV and above the 
$t_1$ dependence becomes noticable. Then the squared reaction amplitude is
significantly reduced so that, after integration over the 3-body phase space,
a perfect description of the reaction cross section at higher energies is 
achieved for $\pi$ as well as for $K$ exchange,  in contrast to the 
calculation where ${\cal M}{=}const.$, cf.  the corresponding results 
in Fig.~\ref{erka1}.
 
This observation suggest that a completely differential
treatment of the reaction $pp{\to}pK^+\Lambda$ within the four-dimensional
space of Eq.~(\ref{full}) is not necessary, because in any case the very 
smooth $t_1$-dependence does not provide access to conclusive information 
about the vertex function, the propagator of the exchange particles and 
the form factor for bombarding energies within the COSY regime. 

\section{The Dalitz plot}

\begin{figure}[t]
\vspace*{-5mm}
\centerline{\hspace*{3mm}\psfig{file=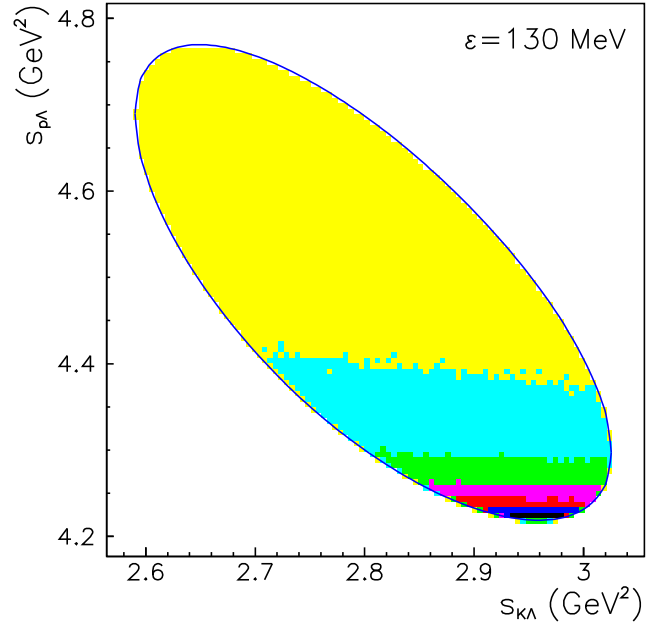,width=9.6cm,height=9.5cm}}
\vspace*{-4mm}
\caption{The Dalitz plot distribution for the reaction $pp{\to}pK^+\Lambda$ 
at the excess energy $\epsilon$=130~MeV as a function of the 
invariant mass squared $s_{K\Lambda}$ and $s_{\Lambda{p}}$. The shown 
result is for the $K$-meson exchange mechanism including the $\Lambda{p}$ FSI. 
The solid contour is the Dalitz plot boundary given by the helicity 
angle $\cos\theta_{31}{=}{\pm}1$ of Eq.~(\ref{helic}).}
\label{erka4c}
\end{figure}

The Dalitz plot for the reaction $pp{\to}pK^+\Lambda$ at the excess
energy $\epsilon$=130~MeV is presented in Fig.~\ref{erka4c}. The
results are based on the $K$-meson exchange mechanism with inclusion 
of the $\Lambda{p}$ FSI.
We consider the excess energy $\epsilon$=130~MeV because we found that 
this is more or less the optimal minimal energy where a separation between 
the FSI effects and the $S_{11}$ resonance is still possible. 
Of course, for the $K$-meson exchange mechanism 
only the structure coming from the $\Lambda{p}$ FSI is detectable in the
Dalitz plot distribution at low $s_{\Lambda{p}}$ and there is no visible
structure due to the $K\Lambda$ subsystem. Recall that for a constant
reaction amplitude ${\cal M}{=}const.$ the distribution is isotropic.
In case of a $K$-meson exchange dominance the experimental Dalitz plot 
should resemble the result shown in Fig.~\ref{erka4c}.

\begin{figure}[t]
\vspace*{-5mm}
\centerline{\hspace*{3mm}\psfig{file=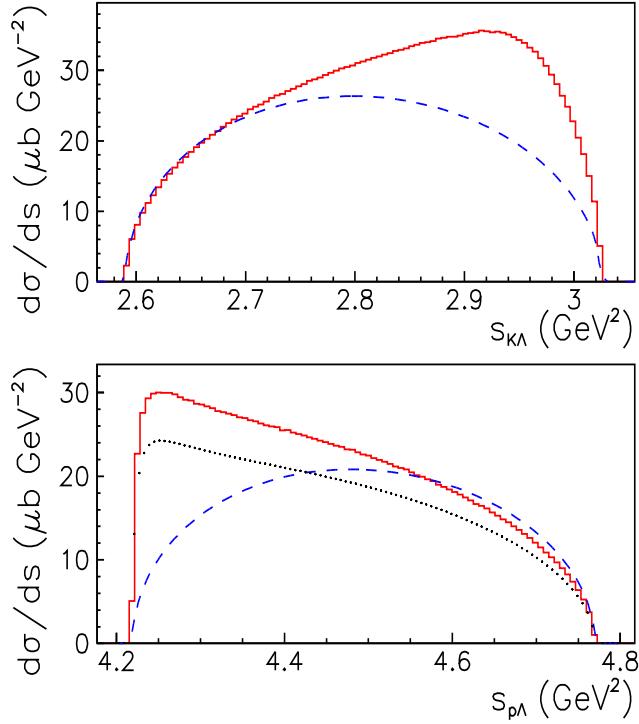,width=9.6cm,height=11.cm}}
\vspace*{-5mm}
\caption{The $s_{K\Lambda}$ and 
$s_{\Lambda{p}}$ invariant mass spectra for the reaction
$pp{\to}pK^+\Lambda$ at $\epsilon$=130~MeV. The solid
histograms show the $K$-meson exchange calculations with inclusion of
the $\Lambda{p}$ FSI. The dashed lines indicate the phase space
distribution given by Eq.~(\ref{phases}), while the dotted line is the
phase space distribution multiplied by the FSI amplitude 
$|{\cal A}_{\Lambda{p}}|^2$ from Eq.~(\ref{jost1}). The dashed and dotted
lines are shown in arbitrary normalization.}  
\label{erka6a}
\end{figure}

The projections of the Dalitz, {\it i.e.} the squared invariant mass
spectra in the $K\Lambda$ and $\Lambda{p}$ subsystems, are shown in 
Fig.~\ref{erka6a}. The solid histograms are calculations for the
$K$-meson exchange mechanism including the $\Lambda{p}$ FSI given by
Eq.~(\ref{jost1}). The dashed lines indicate the phase space
distribution, which results from the integration of the Dalitz plot
of Eq.~(\ref{dalitz}) over one of the invariant mass squared. For
the $\Lambda{p}$ subsystem the squared invariant mass spectrum is given by 
\begin{eqnarray}
\frac{d\sigma}{ds_{\Lambda{p}}}{=}
\frac{\lambda^{1/2}(s,s_{\Lambda{p}},m_K^2)\,
\lambda^{1/2}(s_{\Lambda{p}},m_\Lambda^2,m_p^2)}
{2^8\pi^3\, s \, \sqrt{s^2-4sm_N^2}\, s_{\Lambda{p}}}\,\, |{\cal M}|^2,
\label{phases}
\end{eqnarray}
where $|{\cal M}|^2{=}const.$. The $K\Lambda$ distribution can be easily
obtained in a similar way. The phase space distributions in
Fig.~\ref{erka6a} are arbitrarily normalized. 
As compared to the phase space the result for the $K$-meson exchange
mechanism indicates an enhancement at low $\Lambda{p}$ masses.
At the same time the $K\Lambda$ distribution 
is enhanced at large masses, which results from the kinematic
reflection. We should emphasize, however, that
the enhancement comes practically only from the $\Lambda{p}$ FSI. 

\begin{figure}[b]
\vspace*{-8mm}
\centerline{\hspace*{3mm}\psfig{file=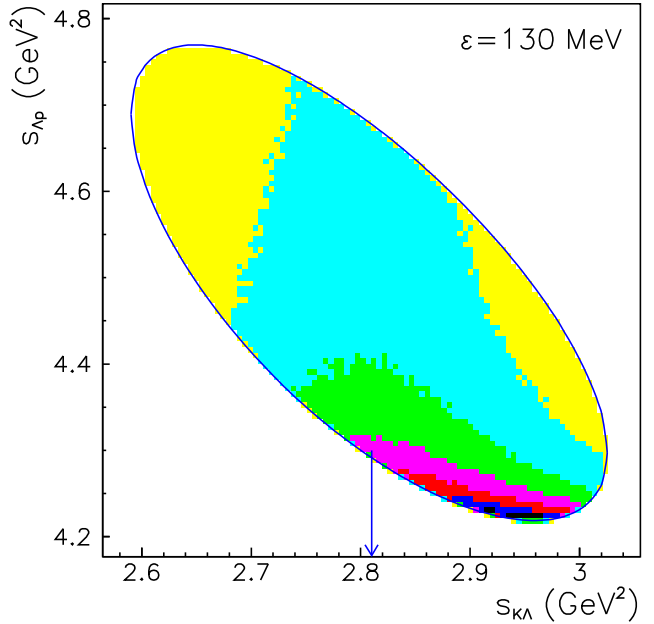,width=9.6cm,height=9.5cm}}
\vspace*{-4mm}
\caption{The Dalitz plot distribution for the reaction $pp{\to}pK^+\Lambda$ 
at $\epsilon$=130~MeV as a function of the invariant mass squared 
$s_{K\Lambda}$ and $s_{\Lambda{p}}$. The shown result is for
the $\pi$-exchange mechanism with excitation of the $S_{11}$(1650) 
resonance and includes also the $\Lambda{p}$ FSI. 
The solid contour is 
the Dalitz plot boundary given by Eq.~(\ref{helic}). The arrow indicates
the square of the $S_{11}$(1650) resonance mass.
}
\label{erka4}
\end{figure}

\begin{figure}[t]
\vspace*{-5mm}
\centerline{\hspace*{3mm}\psfig{file=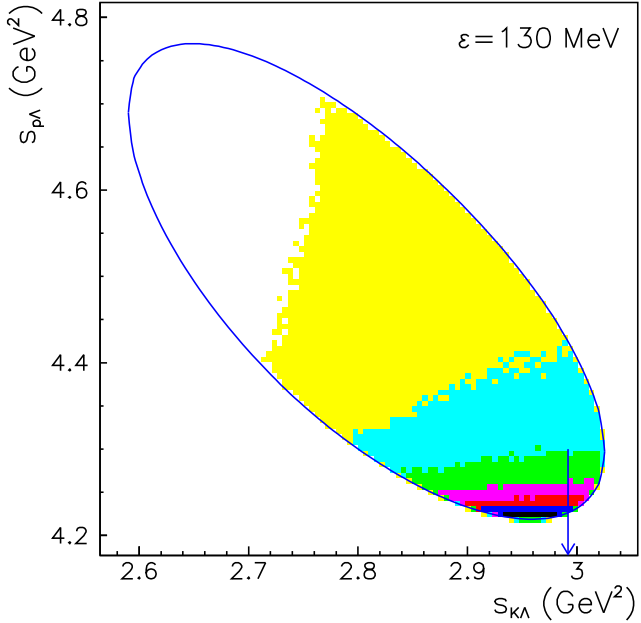,width=9.6cm,height=9.5cm}}
\vspace*{-4mm}
\caption{The Dalitz plot distribution for the reaction $pp{\to}pK^+\Lambda$ 
at $\epsilon$=130~MeV as a function of the invariant mass squared 
$s_{K\Lambda}$ and $s_{\Lambda{p}}$. The shown result is for
the $\pi$-exchange mechanism with excitation of the $P_{11}$(1710) 
resonance and includes also the $\Lambda{p}$ FSI. 
The solid contour is the Dalitz plot boundary given by 
Eq.~(\ref{helic}). The arrow indicates
the squared of the $P_{11}$(1710) resonance mass.}
\label{erka4a}
\end{figure}

Let us now compare the full $K$-meson
exchange calculation with the simple FSI factorization approach given by
the product of the phase space distribution from Eq.~(\ref{phases}) and
the $\Lambda{p}$ FSI amplitude $|{\cal A}_{\Lambda{p}}|^2$ of
Eq.~(\ref{jost1}). Corresponding results are shown by the dotted line in 
Fig.~\ref{erka6a}. The latter was slightly renormalized in order to make it 
optically distinguishable from the (solid) histogram. One can see that the 
$\Lambda{p}$ spectrum obtained by factorization of the FSI and phase space 
practically coincides with the full calculation. This demonstrates that in 
case of an almost constant 
reaction amplitude the $\Lambda{p}$ distribution can be savely used 
for the evaluation of the hyperon-nucleon scattering parameters, as was done
in Refs.~\cite{Hinterberger,Gasparyan}. Note, however, that there is still
a theoretical uncertainty involved in such an evaluation depending strongly
on the method applied~\cite{Gasparyan1}. 
In any case it is clear that, if $K$-meson exchange dominates the 
$pp{\to}pK^+\Lambda$ reaction, then this could be unambigously 
deduced from the Dalitz plot.

Now we turn to the $\pi$-meson exchange mechanism and the excitation 
of baryonic resonances in the $\pi{N}{\to}K\Lambda$ reaction.
The resonances can be recognized by a Breit-Wigner-type shape in the
$\Lambda{K}$ invariant mass spectrum and by the angular dependence of 
the $S_{11}$(1650) and $P_{11}(1710)$ decay products,
which is determined by the resonance spin and production
mechanism~\cite{Byckling}. But one needs to distinguish the 
resonances in the $\Lambda{K}$ subsystem from the $\Lambda{p}$ FSI, 
because the latter mimics a resonance-like 
structure in the $\Lambda{p}$ subsystem with a pole at
$m_\Lambda{+}m_N$ and width given roughly by the $\Lambda{p}$ scattering
parameters. Therefore, it is necessary to consider the complete Dalitz plot
distribution. However, if the baryonic resonances and the FSI overlap we
should return to the completely differential treatment in the 
four-dimensional space given by Eq.~(\ref{full}), {\it i.e.} consider 
the $t_1$ as well as the $t_2$ invariants. 

The Dalitz plot distribution for the $\pi$-meson exchange mechanism,
with excitation of the $S_{11}$(1650) resonance and inclusion of the 
$\Lambda{p}$ FSI, is shown in Fig.~\ref{erka4}. The distribution
substantially differs from the result obtained for the $K$-meson 
exchange scenario. Specifically, the influence of the resonance can be clearly
seen. The arrow in Fig.~\ref{erka4} indicates the resonance position, 
{\it i.e} the square of the resonance mass. A sufficiently large excess
energy like $\epsilon$=130 MeV
allows to separate the effects due to the $S_{11}$(1650) resonance 
and the $\Lambda{p}$ FSI, which is important for the data evaluation. 
The situation is different for $\pi$-meson exchange and $P_{11}$(1710) excitation, 
shown in Fig.~\ref{erka4a}. Here the signal of the $P_{11}$ resonance overlaps 
with the $\Lambda{p}$ FSI. In principle, even
in this case the Dalitz plot might be sufficient to reconstruct the
resonance contribution but it would be more promising to perform a combined 
Dalitz plot and $t_2$ or Jackson-angle analysis. Of course, such
an analysis requires large experimental statistics. 

\begin{figure}[b]
\vspace*{-8mm}
\centerline{\hspace*{3mm}\psfig{file=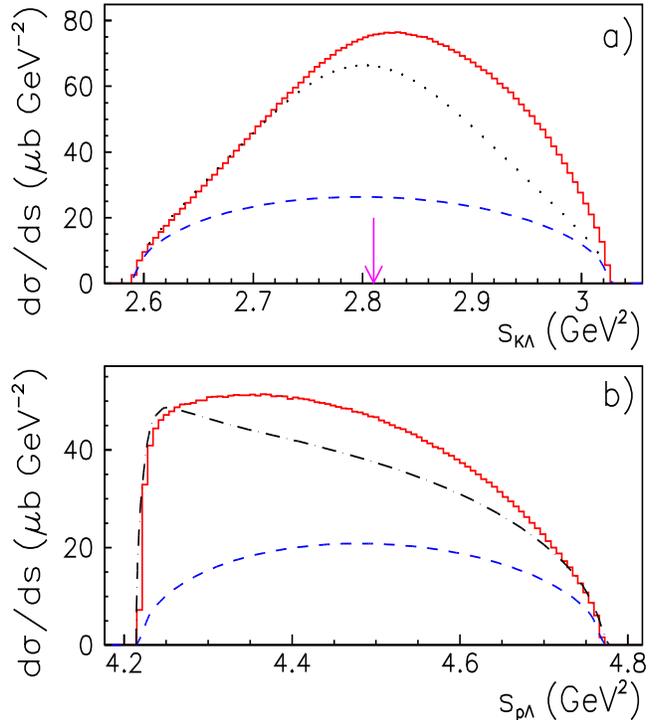,width=9.6cm,height=11.cm}}
\vspace*{-5mm}
\caption{The $s_{K\Lambda}$ and 
$s_{\Lambda{p}}$ squared invariant mass spectra for the reaction 
$pp{\to}pK^+\Lambda$ at $\epsilon$=130~MeV. The solid
histograms show the $\pi$-meson exchange calculation with inclusion
of the $S_{11}$(1650) resonance and the $\Lambda{p}$ FSI.
The dashed lines indicate the phase space
distribution given by Eq.~(\ref{phases}). The dotted line in a) 
is the phase space distribution multiplied by the squared $S_{11}$
resonant amplitude given by Eq.~(\ref{reson}). The dashed-dotted line 
in b) is the
phase space distribution multiplied by the squared FSI amplitude 
$|{\cal A}_{\Lambda{p}}|^2$ from Eq.~(\ref{jost1}). The dashed and dotted
lines are shown with arbitrary normalization. The arrow indicate the 
squared mass of the $S_{11}$(1650) resonance.}  
\label{erka6}
\end{figure}

Since there are now structures in the $K\Lambda$ as well as 
$\Lambda{p}$ subsystems one might expect a substantial distortion of the
Dalitz-plot projections. This issue is addressed in Fig.~\ref{erka6} 
where we show the $s_{K\Lambda}$ and $s_{\Lambda{p}}$ invariant mass spectra. 
The solid histograms are the full results
for the $\pi$-meson exchange mechanism, with excitation of the 
$S_{11}$(1650) resonance and including the $\Lambda{p}$ FSI, 
while the dashed lines indicate the
phase space distributions given by Eq.~(\ref{phases}). 
The dotted line in Fig.~\ref{erka6}b) corresponds to 
the phase space distribution multiplied by the FSI amplitude,
$|{\cal A}_{\Lambda{p}}|^2$, from Eq.~(\ref{jost1}). 
Obviously, and opposite to the $K$-meson exchange scenario
discussed above, now
the factorization in terms of the $\Lambda{p}$ FSI and the phase space 
deviates significantly from the full calculation. The presence of 
of the $S_{11}$(1650) resonance changes the
$\Lambda{p}$ invariant mass spectra. This observation should be kept in
mind when analyzing the invariant mass spectra given by the
projection of the Dalitz plot distribution with the aim to extract 
the $\Lambda p$ effective-range parameters from the FSI. 
A priori the full structure of the reaction amplitude and the effects 
due to possible kinematic reflections in the different final subsystems 
are not known. Thus, one should rather consider slices of the 
Dalitz plot than projections for the aforementioned analysis in order
to be on the save side - though this again requires larger 
experimental statistics.  

The dash-dotted line in Fig.~\ref{erka6}a) is
the phase-space distribution multiplied by the square of the $S_{11}$(1650)
amplitude, cf. Eq.~(\ref{reson}). Again this result differs significantly
from the full $\pi$-meson exchange calculation because of the
kinematic reflection of the $\Lambda{p}$ FSI. Indeed, the enhancement
with respect to the results obtained by factorization at large $s_{K\Lambda}$
stems entirely from the FSI. 

\section{Angular correlations}
As was discussed in Sect. 2 the angular correlations are given by
Eq.~(\ref{jackson}) in the Jackson frame and by Eq.~(\ref{helic}) in 
the helicity frame. While the former angular spectra, {\it i.e.} the Jackson
and Treiman-Yang angular distributions, contain information relevant
for the partial wave decomposition of the reaction amplitude, the
helicity angle is entirely given by kinematics and has no direct
connection with the partial waves amplitudes. 

With regard to the dependence of the reaction amplitude on the $t_1$ 
invariant we concluded already in Sect. 7 that it should be
rather smooth for $\epsilon{<}$200~MeV for $\pi$ as well as for $K$
exchange. Therefore, in this energy range the $t_1$ dependence is not 
a good tool to distinguish between different production mechanism.

In \, case of \, the \, $K$-meson exchange mechanism
the $t_2$ -dependence is related to the $KN$ scattering
amplitude. Since $K^+p$ elastic scattering is dominated by the
$s$-wave one would expect an isotropic distribution of the Jackson angle
of Eq.~(\ref{jackson}). However, the situation should be very different
for the $pn\to pK^0\Lambda$ reaction which involves the $K^+n{\to}K^0p$
subprocess. Due to the strong angular dependence of the charge exchange
amplitude, which originates from the isospin $I{=}0$ component,
the Jackson-angle distribution should exhibit 
$p$-wave contributions~\cite{Sibirtsev4}.  

\begin{figure}[t]
\vspace*{-5mm}
\centerline{\hspace*{5mm}\psfig{file=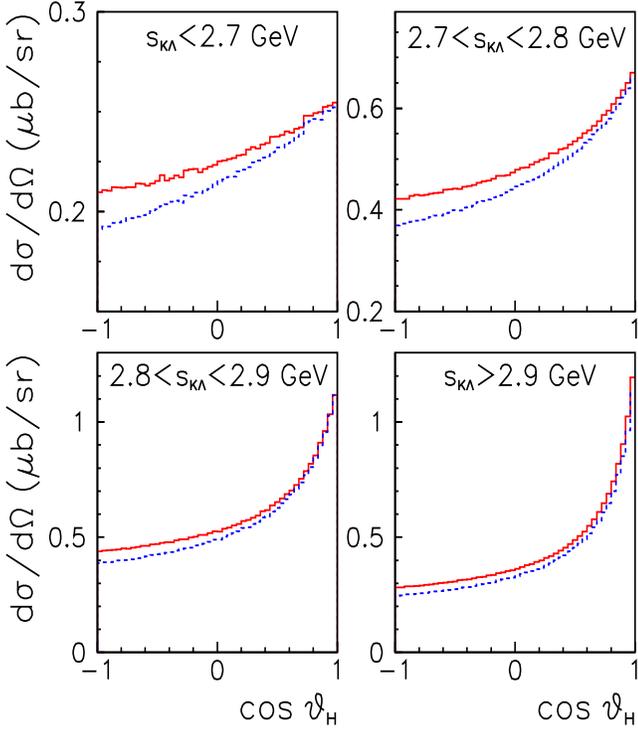,width=9.6cm,height=11.cm}}
\vspace*{-5mm}
\caption{The $\Lambda{p}$ helicity angle spectra at different squared
invariant masses of the $K\Lambda$ subsystem, $s_{K\Lambda}$,
for the reaction $pp{\to}pK^+\Lambda$ at $\epsilon$=130~MeV. The solid
histograms show the $\pi$-meson exchange calculation with inclusion
of the $S_{11}$(1650) resonance and the $\Lambda{p}$ FSI.
The dashed histograms are corresponding results for the $K$-meson
exchange mechanism. 
}  
\label{erka5}
\end{figure}

For the $\pi$-meson exchange mechanism 
the $t_2$ distribution contains the angular dependence of the
$\pi{N}{\to}K\Lambda$ transition amplitude and can be converted 
by Eq.~(\ref{jackson}) to the angular spectrum in the Jackson frame.
If the $S_{11}$(1650) resonance dominates the reaction the Jackson angle
distribution is isotropic, ${\it i.e.}$ similar to that resulting from
the $K^+$-meson exchange scenario. 
If the \, $\pi{N}{\to}K\Lambda$ \, amplitude is \, given entirely by
the $P_{11}(1710)$ resonance, the Jackson angle distribution is again
isotropic, 
which is obvious from Eqs.~(\ref{pwa7a}) and (\ref{pwa7}). 
Finally, if both $S_{11}$ and $P_{11}$ resonances contribute to the 
$pp{\to}pK^+\Lambda$ reaction 
then the Jackson angle distribution would show the interference
between the $s$ and $p$ waves given by Eqs.~(\ref{inva1}) and (\ref{pwa}). 
A proper analysis of the angular distribution would then allow to 
extract the relative contributions of these resonances. 

The angular distribution in the helicity frame just give the
projection of the Dalitz plot as a function of the squared invariant mass
of a particular subsystem while the squared invariant 
mass of the other subsystem is fixed. 
For the reaction $pp{\to}pK^+\Lambda$ one can study the $\Lambda{p}$
invariant mass spectra at a fixed or partially integrated
squared invariant mass of the $K\Lambda$ subsystem. Since $s_{K\Lambda}$
is fixed one can transform, by Eq.~(\ref{helic}), the $s_{\Lambda{p}}$
distribution to the helicity angle distribution. That allows to present
the data in a more convenient way because \hbox{-1${\le}\cos\theta_{31}{\le}1$}. 
But one should remember that the helicity
angle distribution is just a slice of the Dalitz plot and it does not
contain more information than the Dalitz plot itself. We discuss the
usefulness of the helicity angle spectra now.

\begin{figure}[b]
\vspace*{-9mm}
\centerline{\hspace*{5mm}\psfig{file=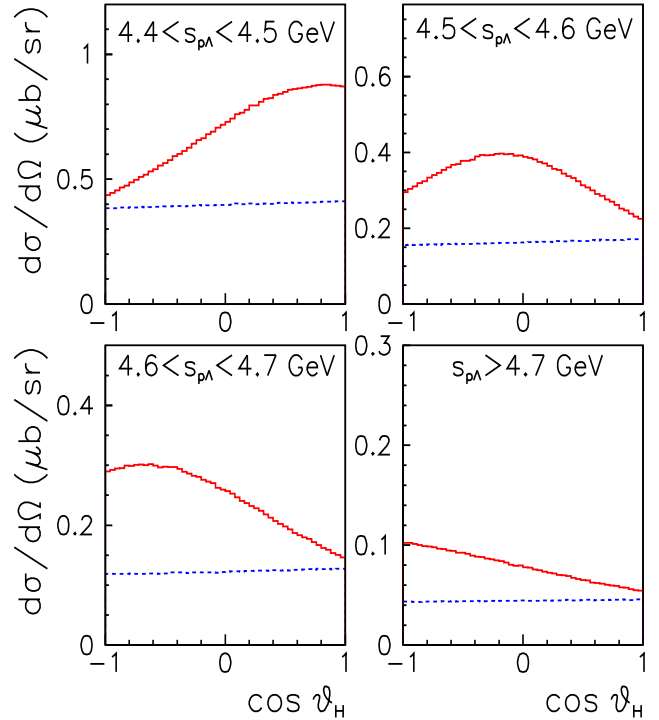,width=9.6cm,height=11.cm}}
\vspace*{-5mm}
\caption{The $K\Lambda$ helicity angle spectra at different squared
invariant masses of the $\Lambda{p}$ subsystem, $s_{\Lambda{p}}$,
for the reaction $pp{\to}pK^+\Lambda$ at $\epsilon$=130~MeV. The solid
histograms show the $\pi$-meson exchange calculations with inclusion
of the $S_{11}$(1650) resonance and the $\Lambda{p}$ FSI. 
The dashed histograms are corresponding results for the $K$-meson
exchange mechanism. 
}
\label{erka7a}
\end{figure}

Fig.~\ref{erka5} shows the $\Lambda{p}$ helicity-angle distribution for
different intervals of the squared invariant mass of the $K\Lambda$
subsystem, for the reaction $pp{\to}pK^+\Lambda$ at
$\epsilon$=130~MeV. Note that according to Eq.~(\ref{helic}) the maximal
$\Lambda{p}$ mass corresponds to forward helicity angles.
The solid histograms are results obtained
for the $\pi$-meson exchange scenario with $S_{11}$(1650) excitation 
and with $\Lambda{p}$ FSI while the dashed histograms are results
for the $K$-meson exchange mechanism. It is obvious that
these distributions are excellent observables for the 
extraction of the $\Lambda{p}$ effective-range parameters, 
since by performing cuts on $s_{K\Lambda}$ one can strongly reduce the
influence of that part of the Dalitz plot which is distorted by the
resonance. Of course, for a separation of the singlet and triplet
parameters corresponding spin-dependent experiments need to be
performed and one should apply reliable extraction methods like the one
advocated in Ref. \cite{Gasparyan1}, based on dispersion theory.

Fig.~\ref{erka7a} shows the $K\Lambda$ helicity-angle distribution 
for different intervals of the squared invariant mass of the $\Lambda{p}$
subsystem. Again the calculations were done at $\epsilon$=130~MeV 
and for the $\pi$-meson exchange mechanism with the $S_{11}$ resonance 
and the $\Lambda{p}$ FSI (solid histograms) and also for the
$K$-meson exchange mechanims (dashed histograms).
We explicitly cut the $\Lambda{p}$ FSI region by taking
the distributions only for $s_{\Lambda{p}}{>}$4.4~GeV$^2$. Now the spectra
show the $S_{11}$(1650) resonance structure whereas they don't show
any structure in case of the $K$-meson exchange. Thus, if there is
any structure these spectra might be fitted by a resonance amplitude, 
i.e. a Breit-Wigner form say, in order to determine the resonance
mass and width. Moreover, the fitting procedure can be applied at
different intervals of $s_{\Lambda{p}}$ following Eq.~(\ref{helic}). 
The procedure should provide resonance parameters independently of the range 
$s_{\Lambda{p}}$ if the FSI region is properly cut. Indeed, fitting 
as a test the calculated $K\Lambda$ helicity-angle spectrum at 
4.6${<}s_{\Lambda{p}}{<}$4.7~GeV$^2$ we obtained the 
resonance parameters $M_R$=1.672~GeV and
$\Gamma$=116.6~MeV. These resonance parameters are close to 
those given by Eq.~(\ref{s11}), i.e. those used for the actual 
calculation with the $\pi$-meson exchange mechanism. 
We examined the procedure by allowing an admixture of 
contributions from the $K$-meson exchange mechanism.
It turned out that the extraction of the resonance parameters from 
the helicity-angle distributions yields quite stable results. 

\section{Summary}

We presented a study of the strangeness production reaction 
$pp{\to}pK^+\Lambda$ for the energy range accessible at 
high-luminosity accelerator facilities like COSY. 
All relevant observables of the reaction for unpolarized beam 
and target nucleons are discussed in terms of their dependence 
on the final four independent invariants. 
The reaction
amplitude is constructed by considering the $\pi$ as well as 
$K$-meson exchange production mechanisms and employing 
elementary $KN{\to}KN$ and $\pi{N}{\to}K\Lambda$ transition
amplitudes taken from a microscopic model ($KN$) and a partial wave 
analysis ($\pi N\to K\Lambda$). Effects of the 
$\Lambda{p}$ final state interaction are included too by means of
the so-called Jost-function approach.

Though our analysis utilizes only $\pi$ and $K$-meson exchange 
we would like to emphasize that the results are, in fact, 
more general. All \, predictions \, given \, for the considered 
spin--in\-de\-pendent observables for the $\pi$-meson exchange 
mechanism, say, would be practically the same for any 
other non-strange meson exchange, {\it i.e.} for the $\sigma$, 
$\eta$, $\rho$ mesons. The quantum numbers and the masses of the 
exchange particle are reflected in the dependence on the squared four 
momentum $t_1$ transfered from the initial to the final
nucleon. However, the range of $t_1$ accessible in experiments with
excess energies up to $\epsilon{\approx}$150 MeV, say, is simply too
small for generating any noticeable differences in the studied 
observables. 
Thus, the two production mechanisms examined in the present 
investigation can be considered as representatives of two general
classes of reaction scenarios, namely where either a nonstrange
or strange meson is exchanged in the production process. 
At the same time this means, of course, that unpolarized experiments
within this energy region won't allow to discriminate more
specifically between different production mechanisms. 

We addressed the questions whether the considered observables can 
be used to determine the $\Lambda p$ interaction or to identify resonances 
that couple to the $K\Lambda$ channel. It was found that 
the Dalitz plot and its sliced projections or the helicity-angle spectra 
are indeed useful for extracting specific information on the $\Lambda{p}$ 
interaction and possible baryonic resonances in the $K\Lambda$ subsystem. 
The Jackson-angle distribution is a crucial tool
to study the onset of higher partial waves in the $KN{\to}KN$
and $\pi{N}{\to}K\Lambda$ transition amplitudes. 
Specifically, 
if the reaction is dominated by the $S_{11}$(1650) resonance we expect
zero $\Lambda$-hyperon recoil polarization and an isotropic distribution 
of the Jackson angle. When both $S_{11}$ and $P_{11}$
resonances contribute the Jackson angle distribution should show an
interference as well as recoil polarization. Furthermore, the Dalitz
plot would explicitly indicate the resonance structure in the
$K\Lambda$ system and at sufficiently large excess energy like 
$\epsilon$=130~MeV the $S_{11}$(1650) resonance effects can be well
isolated from the $\Lambda{p}$ FSI.

We proposed to study specifically the $K\Lambda$ helicity-angle spectra 
at large $\Lambda{p}$ masses squared, $s_{\Lambda{p}}{>}4.4$~GeV$^2$, 
(in order to eliminate effects of the $\Lambda p$ FSI) for a 
reliable determination of ($S_{11}$(1650)) resonance parameters. 
We also pointed out that the $\Lambda{p}$ effective-range parameters 
could be most reliably extracted from the $\Lambda{p}$ helicity-angle 
spectra with cuts $s_{K\Lambda}{>}$2.9~GeV$^2$. 

The results of our calculations are based on a Monte-Carlo integration 
of the three-body phase space, including the mentioned elementary 
reaction amplitudes for $KN$ and $\pi N\to K\Lambda$,
that involves 10$^6$ sample events. 
To determine resonance parameters from the Dalitz plot and
the $K\Lambda$ helicity-angle spectra it is necessary to 
accumulate a data set with large statistics. 
Only then it is possible to achieve an acceptable confidence level 
for the extracted resonance and ($\Lambda p$) effective-range parameters. 
In this context we would like to point out that presently the 
estimates~\cite{PDG} for the mass and 
width of the $S_{11}$(1650) resonance are rather uncertain:
1640${\le}M_R{\le}$1680~MeV and 145$\le\Gamma_R{\le}$190~MeV, 
respectively. The quoted
$P_{11}(1710)$ resonance parameters are 1680 ${\le}M_R{\le}$
1740~MeV for the mass and  
50$\le\Gamma_R{\le}$250~MeV for width. The uncertainties of the 
decay rates of the resonances to the $K\Lambda$ mode is 3--11\% 
and 5--25\% for the $S_{11}$ and $P_{11}$, respectively.

\subsection*{Acknowledgements}
We would like to thank  W.~Eyrich, C.~Hanhart, E.~Kuhlmann, 
U.-G.~Mei{\ss}ner,   J.~Niskanen,
J.~Ritman, E.~Roderburg, S.~Schadmand and W.~Schroeder for useful
discussions. This work was partially supported by the Deutsche
Forschungsgemeinschaft through funds provided to the SFB/TR 16
``Subnuclear Structure of Matter''. This research is part of the EU
Integrated Infrastructure Initiative Hadron Physics Project under
contract number RII3-CT-2004-506078. A.S. acknowledges support by the
COSY FFE grant No.  41760632 (COSY-085).


\begin{thebibliography}{99}
\bibitem{Ferrari1}
        E. Ferrari, Nuovo Cim. {\bf 15}, 652 (1960).
\bibitem{Yao}
	T. Yao, Phys. Rev. {\bf 125}, 1048 (1962).
\bibitem{Wu}
        J.Q. Wu and C.M. Ko, Nucl. Phys. A {\bf 499}, 810 (1989).
\bibitem{Laget}
        J.M. Laget, Phys. Lett. B {\bf 259},24 (1991).
\bibitem{Deloff}
	A. Deloff, Nucl. Phys. A {\bf 505}, 583 (1989).
\bibitem{Sibirtsev1}
        A. Sibirtsev, Phys. Lett. B {\bf 359}, 29 (1995).
\bibitem{Bunce}
	G. Bunce {\it et al.}, Phys. Rev. Lett {\bf 36}, 1113 (1976).
\bibitem{Pondrom}
        L.G. Pondrom, Phys. Rep. {\bf 122}, 57 (1985).
\bibitem{Lundberg}
	B. Lundberg {\it et al.}, Phys. Rev. D {\bf 40}, 3557 (1989).
\bibitem{Smith}
        A.M. Smith {\it et al.}, Phys. Lett. B {\bf 185}, 209 (1987).
\bibitem{Soffer}
        J. Soffer and N.A. T\"ornqvist, Phys. Rev. Lett. {\bf 68}, 907
	(1992).
\bibitem{Turbiner1}
        A.V. Turbiner, JETP Lett {\bf 22}, 182 (1975).
\bibitem{Turbiner2}
        A.V. Turbiner, Sov, J. Nucl. Phys. {\bf 22}, 551 (1976).
\bibitem{Sibirtsev14}
	 A.A. Sibirtsev, Sov. J. Nucl. Phys. {\bf 55}, 145  (1992).
\bibitem{Balewski1}
	J.T. Balewski {\it et al.}, Phys. Lett. B {\bf 388},
	859 (1996).
\bibitem{Balewski2}
	J.T. Balewski {\it et al.}, Phys. Lett. B {\bf 420},
	211 (1998).
\bibitem{Sewerin}
        S. Sewerin {\it et al.}, Phys. Rev. Lett. {\bf 83}, 682 (1999)
        [arXiv:nucl-ex/9811004].
\bibitem{Moskal}
        P. Moskal {\it et al.}, J. Phys. G {\bf 28}, 1777 (2002)
	[arXiv:nucl-ex/0201016].
\bibitem{Wolke}
        M. Wolke {\it et al.}, Nucl. Phys. A {\bf 721}, 683 (2003)
        [arXiv:nucl-ex/0302013].
\bibitem{Kowina1}
        P. Kowina  {\it et al.}, Eur. Phys. J. A {\bf 18}, 351 (2003)
	[arXiv:nucl-ex/0302014].
\bibitem{Kowina2}
	P. Kowina {\it et al.}, Eur. Phys. J. A {\bf 22}, 293 (2004)
	[arXiv:nucl-ex/0402008].
\bibitem{Rozek}
        T. Rozek {\it et al.}, Int. J. Mod. Phys. A {\bf 20},
	625 (2005) [arXiv:nucl-ex/0407024].
\bibitem{Brandt}
	H. Brandt {\it et al.}, Phys. Lett. B {\bf 420}, 217 (1988).
\bibitem{Tsushima1}
        K. Tsushima, A. Sibirtsev, A.W. Thomas, Phys. Lett. B 
	{\bf 390}, 29 (1997) [arXiv:nucl-th/9608029].
\bibitem{Sibirtsev6}
         A. Sibirtsev, K. Tsushima, A.W. Thomas, Phys. Lett. B 
	{\bf 421}, 59 (1998) [arXiv:nucl-th/9711028]. 
\bibitem{Tsushima2}
        K. Tsushima, A. Sibirtsev, A.W. Thomas, Phys. Rev. C 
	{\bf 59}, 369 (1999) [arXiv:nucl-th/9801063].
\bibitem{Faldt}
        G. F\"aldt and C. Wilkin, Z. Phys. A {\bf 357}, 241
	(1997) [arXiv:nucl-th/9612019].
\bibitem{Sibirtsev7}
        A. Sibirtsev, K. Tsushima, W. Cassing, A.W. Thomas,
	arXiv:nucl-th/0004022.
\bibitem{Shyam1}
        R. Shyam, G. Penner, U. Mosel, Phys. Rev. C {\bf 63},
	022202 (2001) [arXiv:nucl-th/0010102]. 
\bibitem{Shyam2}
	 R. Shyam,  Phys. Rev. C {\bf 60}, 055213 (1999)
	[arXiv:nucl-th/9901038]. 
\bibitem{Balewski3}
	J.T. Balewski {\it et al.}, Eur. Phys. J. A {\bf 2},
	99 (1998) [arXiv:nucl-ex/9803008].
\bibitem{Moskal1}
	P. Moskal, M. Wolke, A. Khoukaz, W. Oelert, Prog.
	Part. Nucl. Phys. {\bf 49}, 1 (2002) [arXiv:hep-ph/0208002].
\bibitem{Hinterberger}
        F. Hinterberger and A. Sibirtsev, Eur. Phys. J. A {\bf 21},
	313 (2004) [arXiv:nucl-ex/0402021].
\bibitem{Gasparyan}
        A. Gasparyan, J. Haidenbauer, C. Hanhart, J. Speth,
        Phys. Rev. C {\bf 69}, 034006 (2004) [arXiv:hep-ph/0311116].
\bibitem{Hanhart}
        C. Hanhart, Phys. Rept. {\bf 397}, 155 (2004)
        [arXiv:hep-ph/0311341]. 
\bibitem{Gasparyan1}
	A. Gasparyan, J. Haidenbauer, C. Hanhart, 
	Phys. Rev. C {\bf 72}, 034006 (2005) [arXiv:nucl-th/0506067].
\bibitem{Holzenkamp}
        B. Holzenkamp {\it et al.}, Nucl. Phys. A {\bf 500},
	485 (1989).
\bibitem{Par1} 
	A. Reuber, K. Holinde, J. Speth,
        Nucl. Phys. A {\bf 570}, 543 (1994).
\bibitem{Rijken}
	Th. A. Rijken, V.G.J. Stoks, Y. Yamamoto, Phys. Rev.
	C {\bf 59}, 21 (1999).
\bibitem{Par3} 
        J. Haidenbauer and U.-G. Mei{\ss}ner, 
        Phys. Rev. C {\bf 72}, 044005 (2005) [arXiv:nucl-th/0506019].
\bibitem{Byckling}
        E. Byckling and K. Kajantie, Particle Kinematics, John Willey
	and Sons Pub. (1972).
\bibitem{Dalitz}
        R.H. Dalitz, Phil. Mag. {\bf 44}, 1068 (1953).
\bibitem{Gottfried}
        K. Gottfried and J.D. Jackson, Nuovo Cim. {\bf 33}, 309
	(1964).
\bibitem{Jackson}
	J.D. Jackson, Nuovo Cim. {\bf 34}, 1 (1964).
\bibitem{Treiman}
	S.B. Treiman and C.N. Yang, Phys. Rev. Lett. {\bf 8},
	140 (1962).
\bibitem{Koch}
	W. Koch, Analysis of Scattering and Decay, Gordon and Breach
	Pub. (1968). 
\bibitem{Hoehler}
        G. H\"ohler, Elastic and Charge Exchange Scattering of Elementary
        Particles, Landolt-B\"ornstein, New Series, {\bf 9}, 492 (1983).
\bibitem{Sibirtsev9}
	A. Sibirtsev and W. Cassing, arXiv:nucl-th/9802019.
\bibitem{Gasparian}
        A. Gasparian, J. Haidenbauer, C. Hanhart, L. Kondratyk,  
        J. Speth, Phys. Lett. B {\bf 480}, 273 (2000) [arXiv:nucl-th/9909017].
\bibitem{Buettgen}
        R. B\"uttgen, K. Holinde, A. M\"uller-Groeling, J. Speth,  
        P. Wyborny, Nucl. Phys. A {\bf 506}, 586 (1990).
\bibitem{Hoffmann}
        M. Hoffmann, J.W. Durso, K. Holinde, B.C. Pearce, 
        J. Speth,  Nucl. Phys. A {\bf 593}, 341 (1995).
\bibitem{Haidenbauer}
         J. Haidenbauer and G. Krein, 
         Phys. Rev. C {\bf 68}, 052201 (2003) [arXiv:hep-ph/0309243].
\bibitem{Sibirtsev4}
         A. Sibirtsev, J. Haidenbauer, S. Krewald, Ulf-G. Mei{\ss}ner,
         Phys. Lett. B {\bf 599}, 230 (2004) [arXiv:hep-ph/0405099].
\bibitem{Dolgolenko}
        V.V. Barmin {\it et al.}, Phys. Atom. Nucl. {\bf 66}, 1715 (2003);
        Yad. Fiz. {\bf 66}, 1763 (2003) [arXiv:hep-ex/0304040].
\bibitem{Sibirtsev5}
         A. Sibirtsev, J. Haidenbauer, S. Krewald, Ulf-G. Mei{\ss}ner,
         Eur. Phys. J. A {\bf 23}, 491 (2005).
\bibitem{Berestetsky}
        V.B. Berestetsky and I.Ya. Pomeranchuk, Nucl. Phys. {\bf 22},
        629 (1961).
\bibitem{Ferrari}
        E. Ferrari and F. Selleri, Suppl. Nuovo Cim. {\bf 26}, 451
        (1962).
\bibitem{Sibirtsev2}
         A.A. Sibirtsev, Nucl. Phys. A {\bf 604}, 455 (1996).
\bibitem{Sibirtsev3}
        A. Sibirtsev, W. Cassing, C.M. Ko,  Z. Phys. A {\bf 358},
        101 (1997) [arXiv:nucl-th/9612040].
\bibitem{GaspiN}
        A. Gasparyan, J. Haidenbauer, C. Hanhart, J. Speth,
        Phys. Rev. C {\bf 68}, 045207 (2003) [arXiv:nucl-th/0307072].
\bibitem{Sotona}
        M. Sotona and J. Zofka, Prog. Theor. Phys. {\bf 81}, 160
	(1989).
\bibitem{Krzywicki}
	A. Krzywicki and J. Tran Thanh Van, Phys. Lett. {\bf 30}B,
	185 (1969).
\bibitem{Irving1}
        A.C. Irving, A.D. Martin, C. Michael, Nucl. Phys. 
	B {\bf 32}, 1 (1971).
\bibitem{Hartley}
	B.J. Hartley and G.L. Kane,  Nucl. Phys. 
	B {\bf 57}, 157 (1973).
\bibitem{Irving2}
        A.C. Irving and R.P. Worden, Phys. Rept. {\bf 34}C, 119
	(1977).
\bibitem{Irving3}
        A.C. Irving {\it et al.}, Nucl. Phys. B {\bf 32}, 1 (1971). 
\bibitem{Landolt}
        A. Baldini, V. Flamino, W.G. Moorhead, D.R.O. Morrison, 
	Landolt-B\"ornstein, New Series, {\bf 12} (1988).
\bibitem{Knasel}
        T.M. Knasel {\it et al.}, Phys. Rev. D {\bf 11}, 1 (1975).
\bibitem{Baker}
        R.D. Baker {\it et al.}, Nucl. Phys. B {\bf 141}, 29 (1978).
\bibitem{Goldberger}
	M.L. Goldberger and K.M. Watson, Collision Theory, John Willey
	and Sons (1967).
\bibitem{Sibirtsev10}
	A. Sibirtsev, S. Schneider, C. Elster, J. Haidenbauer,
	S. Krewald, J. Speth, Phys. Rev. C {\bf 65}, 044007 (2002)
	[arXiv:nucl-th/0111086].
\bibitem{Sibirtsev11}
	A. Sibirtsev, S. Schneider, C. Elster, J. Haidenbauer,
	S. Krewald, J. Speth, Phys. Rev. C  {\bf 65}, 067002 (2002)
[arXiv:nucl-th/0203039].
\bibitem{Sibirtsev12}
	A. Sibirtsev, C. Elster, J. Haidenbauer, J. Speth, 
	Phys. Rev. C {\bf 64}, 024006 (2001)
	[arXiv:nucl-th/0104011].
\bibitem{Sibirtsev13}
	A. Sibirtsev, J. Haidenbauer, S. Krewald, U.-G. Mei{\ss}ner,
	Phys. Rev. D {\bf 71}, 054010 (2005)
	[arXiv:nucl-th/0203039].
\bibitem{Watson}
        K.M. Watson, Phys. Rev. {\bf 88}, 1163 (1952).
\bibitem{Migdal}
	A.B. Migdal, JETP {\bf 1}, 2 (1955).
\bibitem{Saxon}
        D.H. Saxon {\it et al.}, Nucl. Phys B {\bf 162}, 522 (1980).
\bibitem{Bell}
	K.W. Bell {\it et al.}, Nucl. Phys. B {\bf 222}, 389 (1983).
\bibitem{PDG}
	S. Eidelman {\it et al.}, Phys. Lett. B {\bf 592}, 1 (2004).
\end{thebibliography}
\end{document}